\documentclass{emulateapj}
\usepackage{amssymb,amsmath}
\usepackage{color,hyperref}
\definecolor{linkcolor}{rgb}{0,0,0.25}
\hypersetup{
  colorlinks=true,        
  linkcolor=linkcolor,    
  citecolor=linkcolor,    
  filecolor=linkcolor,    
  urlcolor=linkcolor      
}
\newcounter{address}
\newcommand{\ie}{i.e.}
\newcommand{\etal}{et al.}
\newcommand{\dd}{\mathrm{d}}
\newcommand{\eg}{e.g.}
\newcommand{\eqnname}{equation}
\newcommand{\Eqnname}{Equation}
\newcommand{\equationname}{\eqnname}

\renewcommand{\figurename}{Figure}

\newcommand{\sectionname}{$\mathsection$}

\renewcommand{\vec}[1]{\ensuremath{\mathbf{#1}}}
\newcommand{\unitvec}[1]{\ensuremath{\mathbf{\hat{#1}}}}
\newcommand{\vecx}{\ensuremath{\vec{x}}}
\newcommand{\vecv}{\ensuremath{\vec{v}}}
\newcommand{\vech}{\ensuremath{\vec{H}}}
\newcommand{\vecj}{\ensuremath{\vec{J}}}

\newcommand{\vecn}{\ensuremath{\vec{n}}}
\newcommand{\veco}{\ensuremath{\vec{\Omega}}}
\newcommand{\veca}{\ensuremath{\boldsymbol\theta}}

\newcommand{\sigv}{\ensuremath{\sigma_v}}

\newcommand{\Myr}{\ensuremath{\,\mathrm{Myr}}}
\newcommand{\Gyr}{\ensuremath{\,\mathrm{Gyr}}}
\newcommand{\kpc}{\ensuremath{\,\mathrm{kpc}}}
\newcommand{\pc}{\ensuremath{\,\mathrm{pc}}}
\newcommand{\kms}{\ensuremath{\,\mathrm{km\ s}^{-1}}}
\newcommand{\msun}{\ensuremath{\,\mathrm{M}_{\odot}}}
\newcommand{\inv}{\ensuremath{^{-1}}}

\newcommand{\apar}{\ensuremath{\theta_\parallel}}
\newcommand{\opar}{\ensuremath{\Omega_\parallel}}
\newcommand{\aperp}{\ensuremath{\veca_\perp}}
\newcommand{\aperpii}{\ensuremath{\theta_{\perp,ii}}}
\newcommand{\operp}{\ensuremath{\veco_\perp}}
\newcommand{\operpii}{\ensuremath{\Omega_{\perp,ii}}}
\newcommand{\ts}{\ensuremath{t_s}}

\newcommand{\vlos}{\ensuremath{V_{\mathrm{los}}}}
\newcommand{\pmll}{\ensuremath{\mu_l}}
\newcommand{\pmbb}{\ensuremath{\mu_b}}

\submitted{}

\begin{document}

\title{Dynamical modeling of tidal streams}
\author{Jo~Bovy$^1$}
\affil{Institute for Advanced Study, Einstein Drive, Princeton, NJ 08540, USA}
\email{Email: bovy@ias.edu}
\altaffiltext{\theaddress}{\label{Hubble}\stepcounter{address} Hubble fellow}

\begin{abstract} 
  I present a new framework for modeling the dynamics of tidal
  streams. The framework consists of simple models for the initial
  action--angle distribution of tidal debris, which can be
  straightforwardly evolved forward in time. Taking advantage of the
  essentially one-dimensional nature of tidal streams, the
  transformation to position--velocity coordinates can be linearized
  and interpolated near a small number of points along the stream,
  thus allowing for efficient computations of a stream's properties in
  observable quantities. I illustrate how to calculate the stream's
  average location (its ``track'') in different coordinate systems,
  how to quickly estimate the dispersion around its track, and how to
  draw mock stream data. As a generative model, this framework allows
  one to compute the full probability distribution function and
  marginalize over or condition it on certain phase--space dimensions
  as well as convolve it with observational uncertainties. This will
  be instrumental in proper data analysis of stream data. In addition
  to providing a computationally-efficient practical tool for modeling
  the dynamics of tidal streams, the action--angle nature of the
  framework helps elucidate how the observed width of the stream
  relates to the velocity dispersion or mass of the progenitor, and
  how the progenitors of ``orphan'' streams could be located.

  The practical usefulness of the proposed framework crucially depends
  on the ability to calculate action--angle variables for any orbit in
  any gravitational potential. A novel method for calculating actions,
  frequencies, and angles in any static potential using a single orbit
  integration is described in an Appendix.
\end{abstract}


\section{Introduction}

Tidal streams hold enormous promise as probes of both the large-scale
structure of the Milky Way (MW) halo's density distribution
\citep[\eg,][]{Johnston99a,Koposov10a} and its small-scale
fluctuations \citep{Carlberg12a}. However, two factors have hampered
the practical use of tidal streams in obtaining constraints on the MW
gravitational potential. First, streams do not trace a single orbit,
which has led to confusion over how to best fit streams and over how
problematic the single-orbit approximation really is
\citep{Eyre11a,Sanders13a}. Second, approaches that go beyond the
single-orbit assumption encounter practical difficulties of
computational cost and observational data quality making many of these
approaches impractical for real, noisy data
\citep{Law05a,Sanders13b,PriceWhelan13a}. Observed gaps in tidal
streams may be due to interactions with dark--matter subhalos
\citep{Yoon11a}, but underdensities could also be created by the
dynamics of stream stars \citep{Kuepper10a}. This hinders our ability
to use underdensities in star counts along a tidal stream in order to
constrain the number of encounters with dark satellites and their
masses \citep{Ngan14a}. Simple, analytic models of how tidal streams
are generated---\emph{generative} models---would greatly benefit both
of these applications.

It has long been clear that the dynamics of tidal streams is most
simply described in terms of action--angle coordinates
\citep{Tremaine99a,Helmi99a}. Once a star has been tidally stripped
from the progenitor, the self-gravity of the stream can be neglected
and the orbital actions of a stream member are conserved while the
angles increase linearly with time. The action--angle structure of a
stream is therefore characterized by strong but straightforward
correlations between the actions and angles of stream members, the
exploitation of which is crucial for using streams to measure the host
gravitational potential \citep{Sanders13b}. The description of a
stream in action--angle coordinates also elucidates the connection
between the orbit and velocity dispersion (or mass) of the progenitor
and the action distribution of the tidal debris \citep{Eyre11a},
allowing simple physical models of the stream to be used and to be
constrained by observational data.

However, streams are observed in position--velocity coordinates and
for action--angle descriptions to be useful, we must be able to
calculate the transformation between these coordinate systems
efficiently. Until now this has required accurate phase--space data
and specialized algorithms that break down for the eccentric orbits on
which tidal-stream progenitors are typically found (that is, radial
and/or vertical actions of similar magnitude as the angular momentum;
\citealt{Sanders12a}). However, even with Gaia \citep{Perryman01a},
all of the phase--space coordinates except for the sky position will
typically have non-negligible uncertainties compared to the intrinsic
dispersion of the stream and in particular the line-of-sight
velocities of the faintest stream members will not be observed for
many stars. Additionally, streams are superimposed on a non-negligible
background of field stars and background contamination cannot be
easily taken into account in any of the current stream-fitting
methods.

In this paper, I present a new method for modeling tidal streams that
fundamentally lives in frequency--angle space, because the stream
distribution function is essentially one-dimensional in this
space. While many of the results of this paper are more generally
valid, the fiducial stream model in this paper consists of a
three-dimensional (close to) Gaussian distribution of frequencies in
the stream, a three-dimensional Gaussian distribution of initial angle
offsets between stream members and the progenitor, and a uniform
distribution of stripping times; the frequency and angle offset
distributions are independent of stripping time. While this model
oversimplifies the real stripping process, I hypothesize that at least
for constraining the gravitational potential with stream data, this
simple model is likely unbiased. For detailed mock stream data, more
elaborate models of the stripping process as a function of time might
be necessary and I discuss how these could easily be incorporated into
the proposed framework.

To evaluate the model in position--velocity coordinates, the
transformation to and from action--angle coordinates is computed using
a novel method for calculating action--angle coordinates presented in
Appendix~\ref{sec:aa}. This transformation is linearized at a small
number of points near the stream that span its length. For streams
originating from progenitors with masses $\lesssim 10^8\msun$, the
separation between the stream track and the progenitor orbit as well
as the internal differences within the stream are small, such that the
linear approximations used here are accurate enough. The linear
approximation allows for fast evaluation of the average stream
location (the stream ``track'') in position--velocity coordinates and
efficient mock data generation. For a uniform distribution of
stripping times, an observed stream member's phase--space probability
distribution function (PDF) can be analytically marginalized over
stripping time, such that using the linear approximation any
evaluation---including marginalization and uncertainty
convolution---of the PDF is extremely rapid. This clears the way for
proper probabilistic inference of the Milky Way halo's gravitational
potential and progenitor properties using stream data for individual
stars.

The framework presented in this paper differs from other approaches
that model streams in action--angle coordinates in a few respects. As
in \citet{Helmi99a}, the model is a generative one in that it gives an
analytic prescription for the initial action--angle distribution of
tidal debris that is then evolved in time to generate a tidal
stream. The difference between these two generative models is that I
propose simple, few-parameter models for the initial action--angle
distributions, that I work out the properties of streams in observable
coordinates in more detail, and that I present and use a general
action--angle transformation that allows stream modeling in general
potentials. The average location of a stream in phase-space can be
computed in the framework presented here and it could be used to
replace orbit fitting; \citet{Varghese11a} discussed a similar method
to model observed streams. The method in this paper differs from that
of \citet{Varghese11a} in that the model is fundamentally located in
action--angle space rather than position--velocity space and in that
the width and length of the stream are related to the offsets between
the stream, the progenitor's orbit, and the average orbit as a
function of stream position. Thus, the present framework does not have
to assume a progenitor velocity dispersion, but can constrain it from
the stream data as well. My approach differs from the framework of
\citet{Johnston98a}, which is also a generative model, in that it
lives in action--angle space rather than energy--angular-momentum
space and that it can be applied to arbitrary time-independent
potentials rather than just spherical potentials close to a
logarithmic potential.

This paper is outlined as follows. In \sectionname~\ref{sec:dynamics}
I briefly summarize the dynamics of tidal streams in terms of
action--angle coordinates. A generative model of a tidal stream in
frequency--angle coordinates is given in
\sectionname~\ref{sec:generative}. \sectionname~\ref{sec:track}
describes how to compute the stream's properties as a function of
angle along the stream, both in action--angle coordinates and in
position--velocity coordinates. \sectionname~\ref{sec:mock} discusses
how to generate mock stream data using the generative model of a tidal
stream and in \sectionname~\ref{sec:pdf} I show how to calculate the
stream PDF for individual stream members as well as its
marginalization and convolution over missing and noisy directions. A
discussion and outlook is presented in
\sectionname~\ref{sec:discussion} and I conclude in
\sectionname~\ref{sec:conclusion}. In \appendixname~\ref{sec:aa} I
review the method of \citet{Fox12a} for calculating actions in any
static gravitational potential and I show how it can be simplified and
extended to compute frequencies and angles.

\section{The dynamics of tidal streams}\label{sec:dynamics}

The dynamics of tidal streams both in position--velocity and
action--angle space has been described previously by various authors
\citep[\eg,][]{Helmi99a,Tremaine99a,Johnston98a,Sanders13a}. I provide
here a brief description of the dynamics of tidal streams that is
relevant for what follows.

In position--velocity space, a tidal stream forms as stars are
stripped from a progenitor cluster or satellite galaxy some time $\ts
> 0$ in the past and then evolve (largely) independently in a galaxy's
host potential, moving away both in position and velocity from the
progenitor object as time goes on. Thus, a star was offset by a small
amount $(\Delta \vecx,\Delta \vecv)$ from the position of the
progenitor $(\vecx^p,\vecv^p)(-\ts)$ at that time. Then the star
orbited under the influence of the same gravitational potential $\Phi$
as the progenitor until the present day $(t=0)$, when it was observed:
$(\vecx,\vecv)(t=0)$ = $\vech_{\ts}(\vecx^p(-\ts)+\Delta
\vecx,\vecv^p(-\ts)+\Delta \vecv)$, where $\vech$ denotes the
Hamiltonian flow from $t=-\ts$ to $t=0$ (see
\citealt{binneytremaine}); this is simply the orbit of the star
between these times. In phase-space this orbit can be calculated by
solving Hamilton's equations: $\dot{\vecx} = \vecv; \dot{\vecv} = -
\dd \Phi / \dd \vecx$.

In action--angle space the dynamics of stream formation is much
simpler. A star received an offset ($\Delta \vecj,\Delta \veca$) at
time $-\ts$ from the action--angle coordinates of the progenitor
$(\vecj^p,\veca^p)$, similar to the small offset ($\Delta \vecx,\Delta
\vecv)$. The offset in the actions corresponds to an offset in the
frequencies $\Delta \veco$; ultimately, this frequency offset is
responsible for the spreading of the stream over long stretches on the
sky. Because both the progenitor's angles and the star's angles
increase linearly with time, albeit at different frequencies, the
difference in the angles also increases linearly in time. Therefore at
time $t=0$ the difference in angles is
\begin{equation}
\Delta \veca(t=0) = \Delta \veca(-\ts) + \Delta \veco \,\ts\,,
\end{equation}
while the difference in actions is constant
\begin{equation}
\Delta \vecj(t=0) = \Delta \vecj(-\ts)\,.
\end{equation}
As the initial offset $\Delta \veca(-\ts)$ in angles is small compared
to the growth $\Delta \veco \,\ts$, we have that $\Delta \veca(t=0)
\approx \Delta \veco \,\ts$.

Thus, in action--angle coordinates, the dynamics of tidal-stream
formation is simplified to linear growth of small initial angle
differences due to small action/frequency differences. The angle
direction $\unitvec{\veco} \equiv \Delta \veco / | \Delta \veco|$
along which a star moves away from the progenitor is related to the
initial action offset by $\Delta \veco \approx (\partial \veco /
\partial \vecj) \Delta \vecj = (\partial^2 H / \partial \vecj \partial
\vecj) \Delta \vecj$, in the limit of small action offsets, because
$\veco = \partial H / \partial \vecj$, where $H$ is the
Hamiltonian. For an approximately one-dimensional stream to form, the
Hessian matrix $(\partial^2 H / \partial \vecj \partial
\vecj)\large|_{\vecj^p}$ has to be dominated by a single large
eigenvalue; in this case the angle difference of stars stripped from
the stream at any time will fall along approximately the same
direction $\unitvec\veco$, and the stream will be essentially one
dimensional. In detail, because the distribution of initial offsets
$\Delta \vecj$ is not isotropic, the direction $\unitvec \veco$ is not
just determined by the potential (through $H$) and the progenitor
orbit $\vecj^p$, but also by the distribution of $\Delta \vecj$, which
is set through a combination of the internal dynamics of the
progenitor and the stripping process.

\section{Generative models of tidal streams}\label{sec:generative}

\subsection{A simulation}\label{sec:sim}

I illustrate the theoretical considerations set forth in this paper by
using a simulated tidal stream. This stream is set up similarly to
that used in \citet{Sanders13b}. The stream is constructed by running
an $N$-body simulation of a King cluster \citep{King66a} on an orbit
similar to that of the GD-1 stream \citep{Koposov10a}. Specifically, a
King cluster with a mass of $2\times10^4\msun$, a tidal radius of
$0.07\kpc$, and a ratio of the central potential to the velocity
dispersion squared of $2.0$ is sampled using $10^4$ particles. The
cluster is evolved with self-gravity in an external logarithmic host
potential with $V_c = 220\kms$ and a flattening $q=0.9$ ($\Phi = V_c^2
\ln [R^2 + z^2/q^2]$) using the gyrfalcon code
\citep{Dehnen00a,Dehnen02a} with a softening of $1.5\pc$ in the NEMO
toolkit \citep{Teuben95a}. The cluster is evolved for $5.011\Gyr$
($=5.125\kpc/[\kms]$) from the initial condition $(X,Y,Z) =$
$(-11.63337239$ $,-20.76235661,$ $-10.631736273934635)\kpc$ and
$(V_X,V_Y,V_Z) = $ $(-128.8281653$ $,42.88727925,$ $79.172383882274971)\kms$,
chosen such that the cluster ends up at roughly the observed location
of the GD-1 stream at the end of the simulation
\citep{Koposov10a}. The progenitor cluster's orbit has an eccentricity
of $0.31$, a pericenter of $13.7\kpc$, and a current $r=14.4\kpc$,
such that the cluster is close to pericenter at the end of the
simulation. The actions of the progenitor are $(J_R,L_Z,J_Z) =
(288.5,-3173.7,897.6)\kpc\kms$ and its orbital frequencies are
$(\Omega_R,\Omega_\phi,\Omega_Z) = (15.7,-10.8,11.9)\Gyr\inv$.

\begin{figure}[t!]
  \includegraphics[width=0.48\textwidth,clip=]{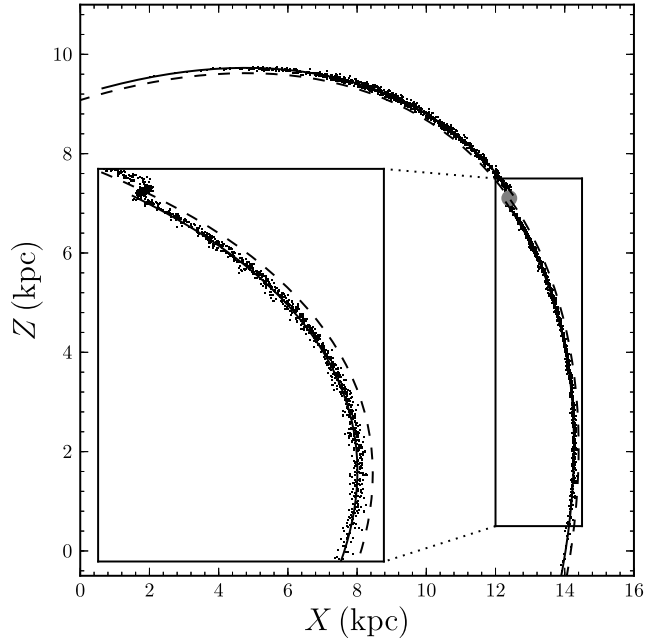}
  \caption{Mock tidal stream simulated as described in
    \sectionname~\ref{sec:sim} after $5.011\Gyr$ of evolution in
    Galactocentric $X$ and $Z$ coordinates. The gray dot shows the
    current position of the progenitor cluster and the dashed line
    shows its orbit. The solid line gives the stream track, calculated
    using the method described in \sectionname~\ref{sec:track}. The
    inset shows a zoomed-in version of part of the leading stream
    (stretched in $X$ and $Z$) that more clearly shows the offset
    between the stream track and the progenitor
    orbit.}\label{fig:gd1_xz}
\end{figure}

\begin{figure*}[t!]
  \includegraphics[width=0.32\textwidth,clip=]{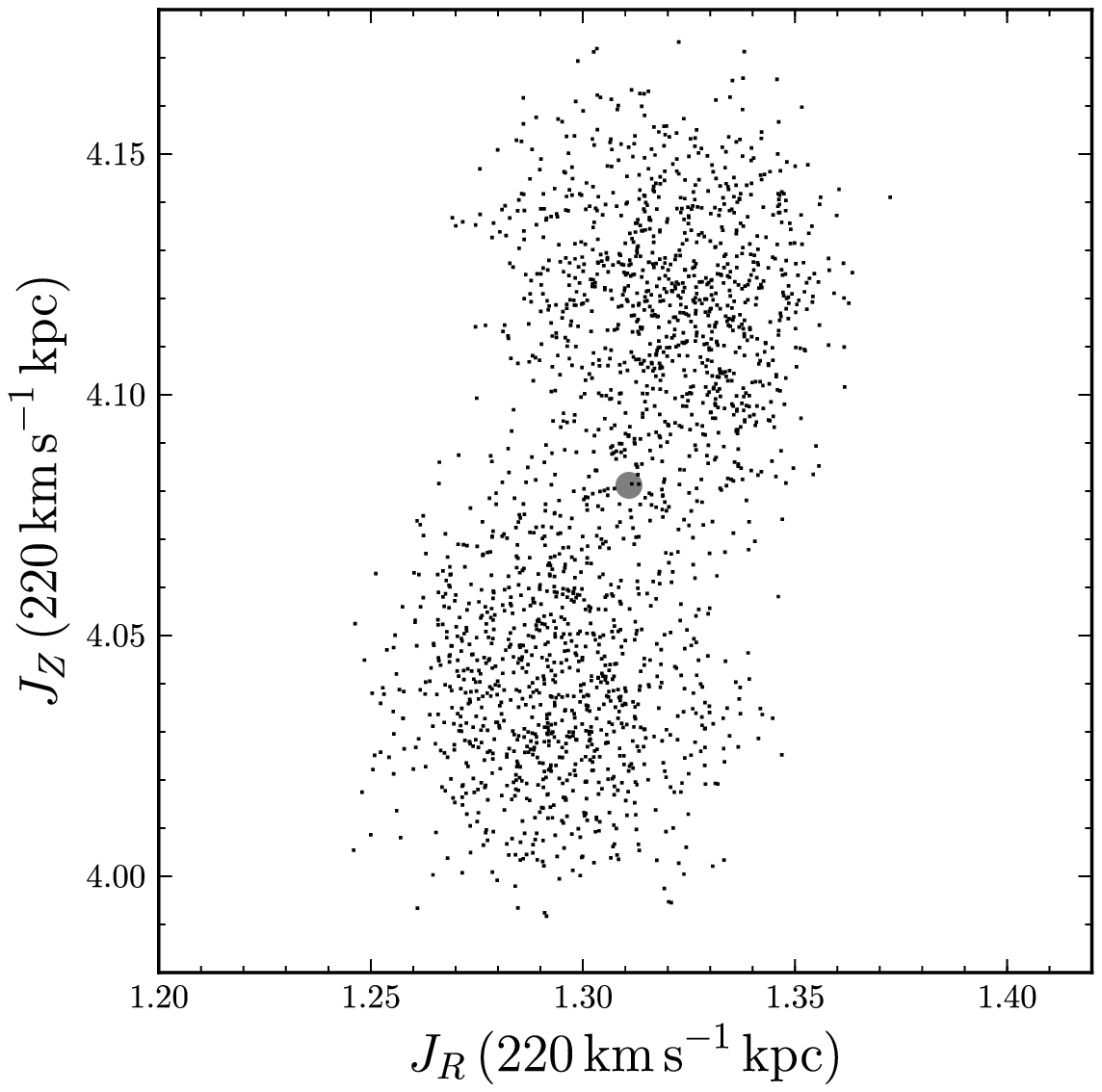}
  \includegraphics[width=0.32\textwidth,clip=]{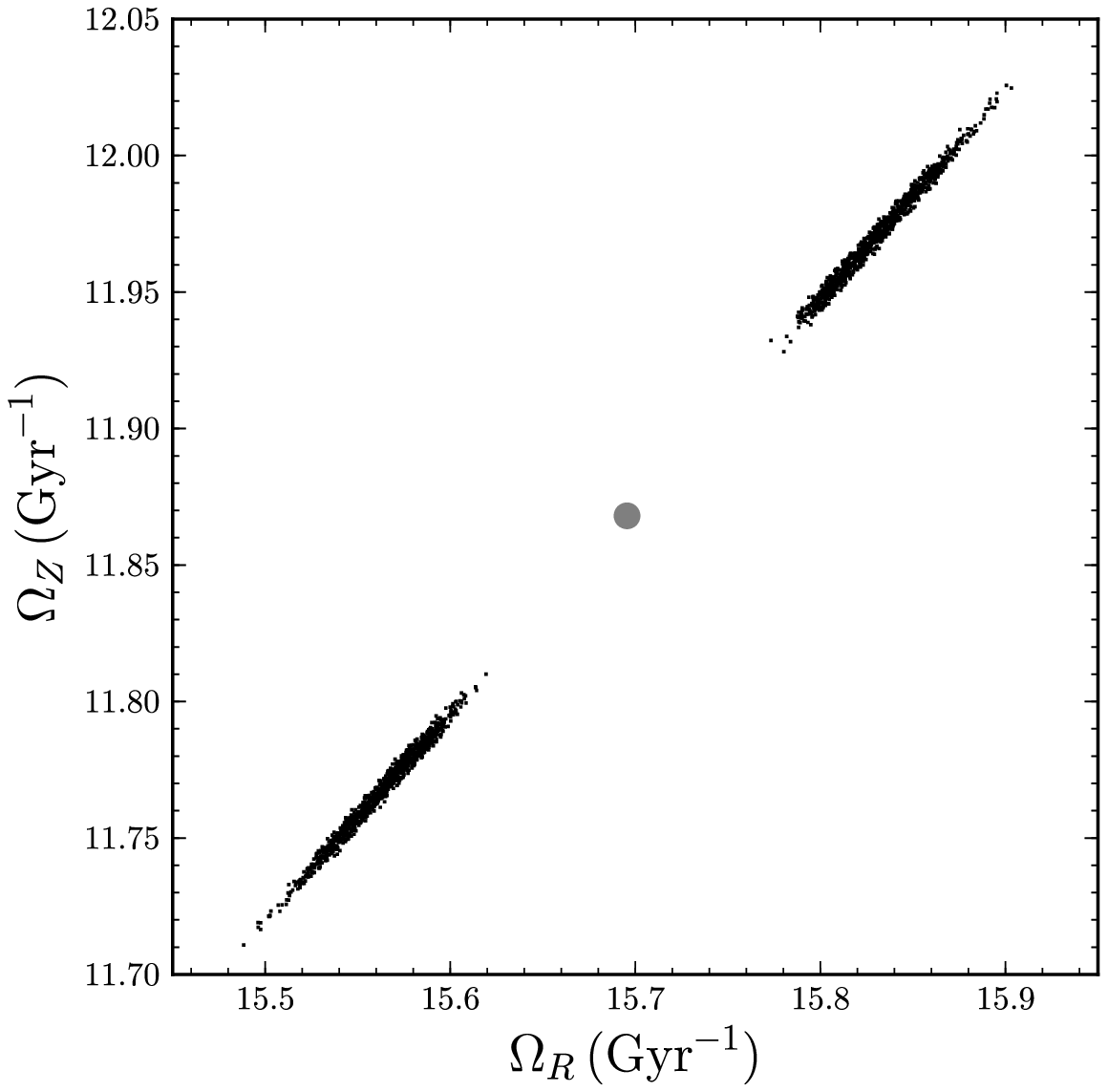}
  \includegraphics[width=0.32\textwidth,clip=]{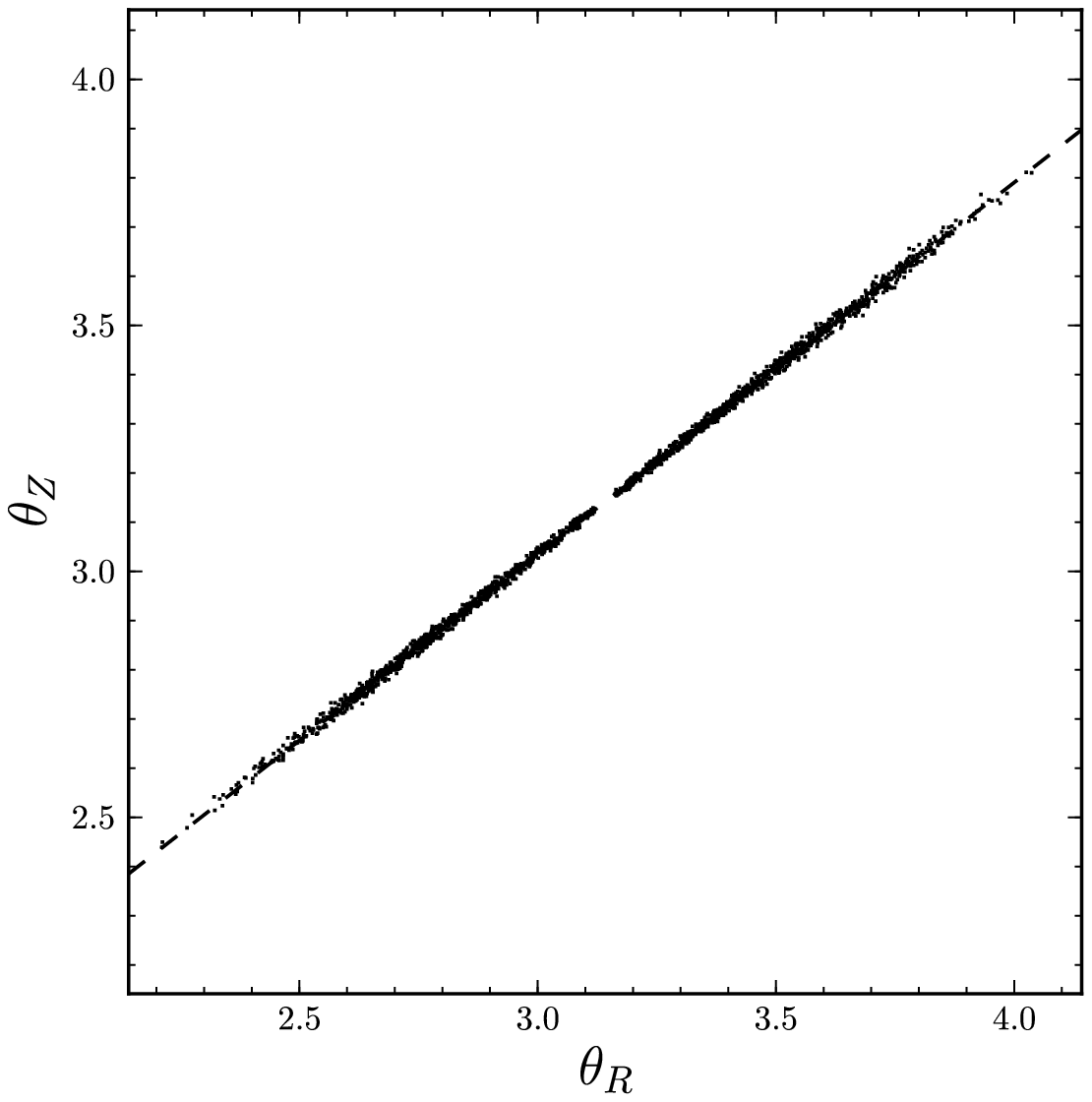}\\
  \includegraphics[width=0.32\textwidth,clip=]{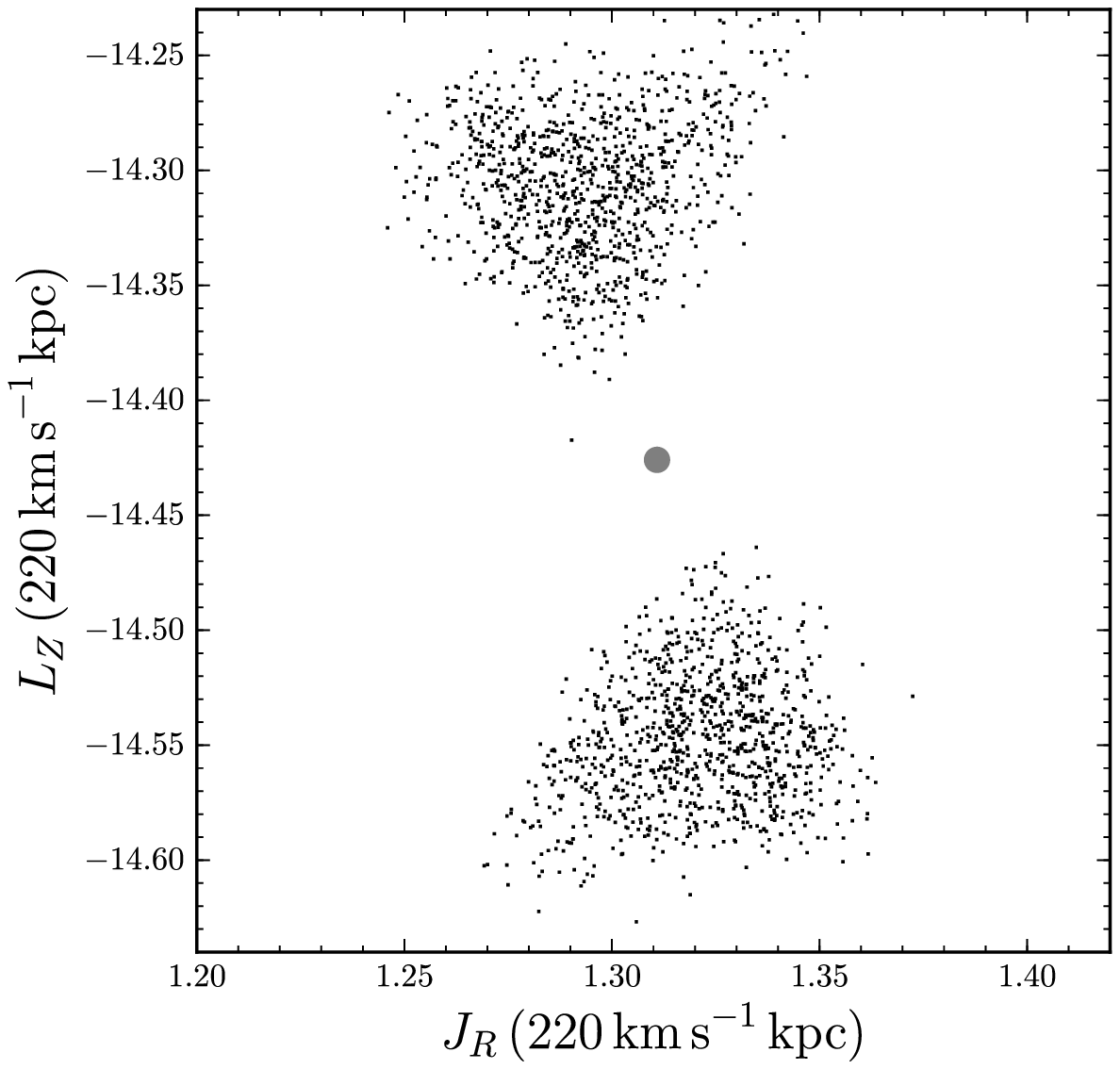}
  \includegraphics[width=0.32\textwidth,clip=]{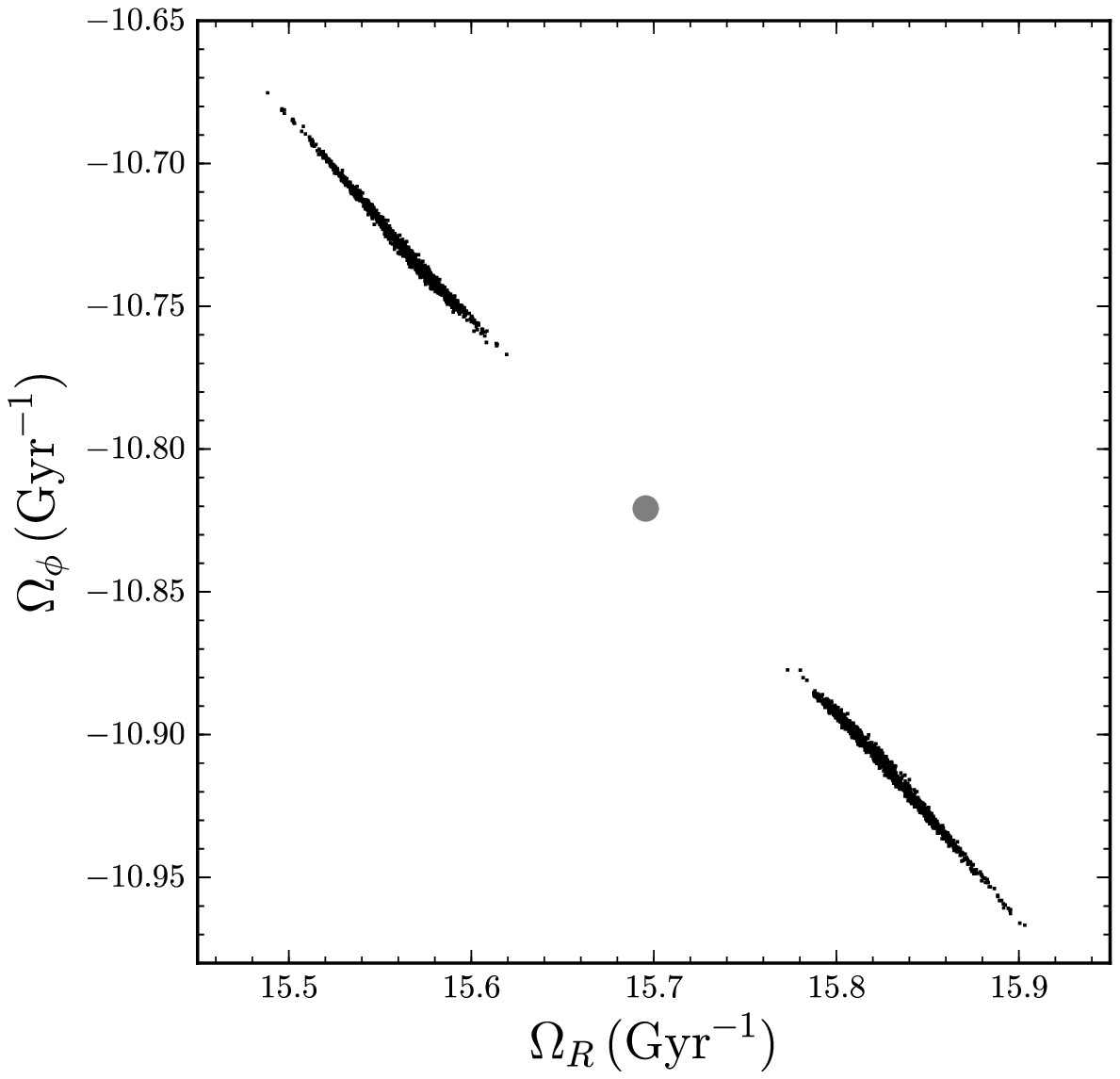}
  \includegraphics[width=0.32\textwidth,clip=]{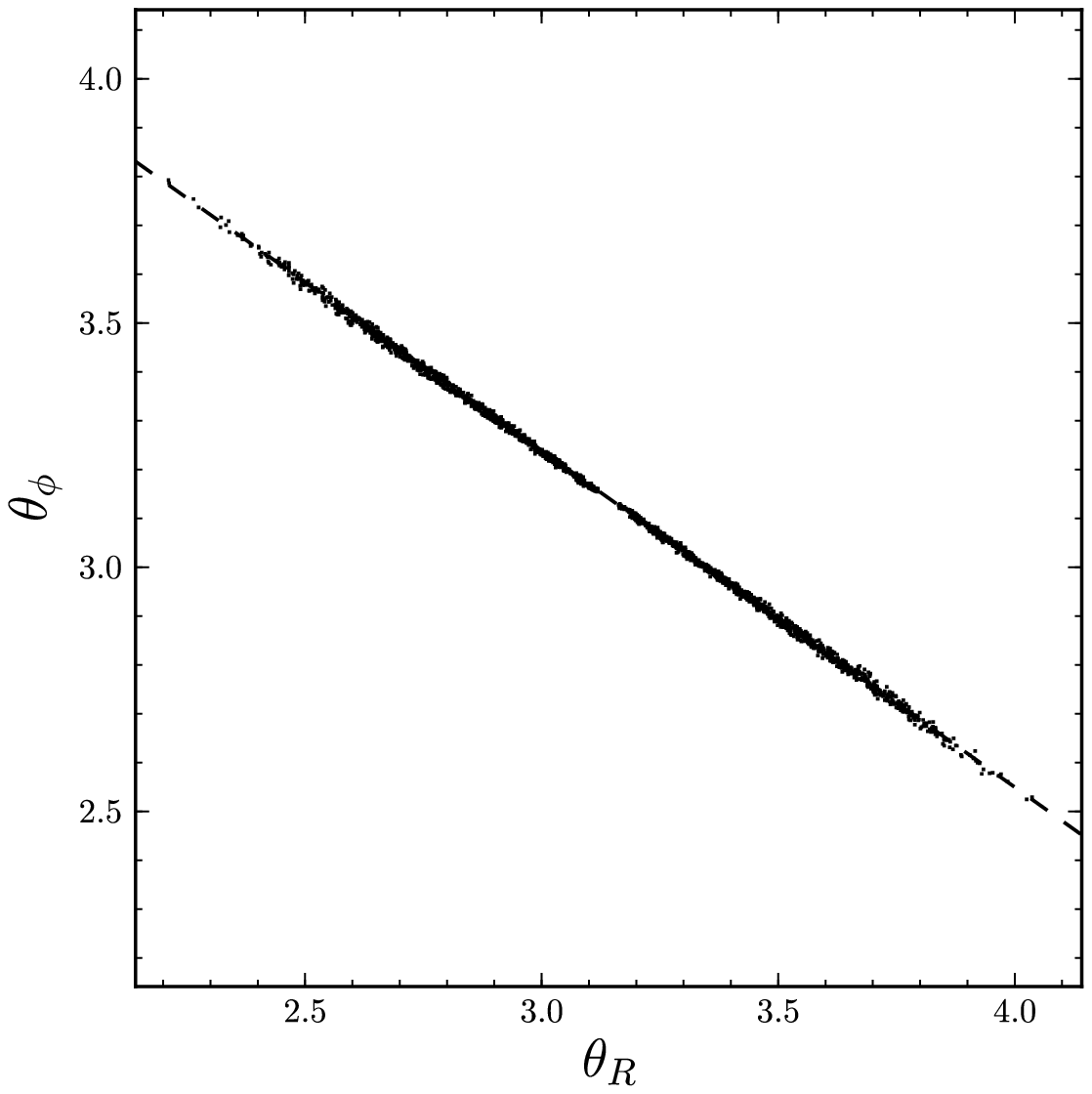}
  \caption{Simulated tidal stream from \sectionname~\ref{sec:sim}
    after $5.011\Gyr$ of evolution in action--frequency--angle
    coordinates. The top row displays projections of the actions
    (left), frequencies (middle), and angles (right) in the $R$ and
    $Z$ directions; the bottom panels shows the same for $R$ and
    $\phi$. While the action distribution has a quite complicated
    form, the frequency distribution is close to one-dimensional and
    the leading and trailing parts of the stream clearly separate (the
    two clouds in action--action and frequency--frequency correspond
    to the leading and trailing arm). In the right panels, the angles
    of the progenitor have each been shifted to $\pi$ and the dashed
    lines show the direction of the progenitor's frequency vector. The
    misalignment between the progenitor's frequency vector and the
    mean frequency-offset vector of the debris is
    $0.18^\circ$.}\label{fig:gd1_jt}
\end{figure*}

The final position of the stream in $(X,Z)$ is shown in
\figurename~\ref{fig:gd1_xz}. \figurename~\ref{fig:gd1_jt} shows the
final position of the stream in action--frequency--angle coordinates,
calculated using the algorithm described in
\appendixname~\ref{sec:aa}. As described in
\sectionname~\ref{sec:dynamics}, a one-dimensional tidal stream forms
because even though the action distribution of the tidal debris is not
isotropic or one-dimensional, the Hessian $(\partial^2 H / \partial
\vecj \partial \vecj)\large|_{\vecj^p}$ is so strongly dominated by a
single large eigenvalue (the ratio of the largest to the second
largest eigenvalue is about $30$), that the frequency distribution of
the debris is essentially one dimensional. The angle distribution of
the debris at the end of the simulation is therefore essentially one
dimensional as well.

\subsection{A model in frequency--action space}\label{sec:modeloa}

The action and frequency distributions shown in
\figurename~\ref{fig:gd1_jt} combined with the knowledge that for
\emph{any} tidal stream the Hessian $(\partial^2 H / \partial \vecj
\partial \vecj)\large|_{\vecj^p}$ must be strongly dominated by a
single large eigenvalue leads me to propose that frequency--angle
space is a better coordinate system than action--angle space for
modeling tidal streams. That is, the distribution of actions in a
tidal stream has a complicated structure that is hard to capture using
a simple, analytic form (see also the discussion in
\citealt{Eyre11a}). Especially the ``bow tie'' structure in
$(J_R,L_Z)$ and the overlapping leading- and trailing-arm
distributions in ($J_R,J_Z$) are difficult to model with a simple
distribution function\footnote{For triaxial potentials, the azimuthal
  action $J_\phi \neq L_Z$ is the correct action to use instead of
  $L_Z$. Even though the modeling in this paper also applies to
  triaxial potentials, we denote $J_\phi$ as $L_Z$, because the
  simulation uses an axisymmetric potential.}. The distribution of
frequencies, however, is close-to one dimensional, its direction of
largest variance can be well-modeled as a Gaussian (although below we
will model it slightly differently), and its parameters can all be
easily estimated from the velocity dispersion of the progenitor, the
orbit of the progenitor, and the gravitational potential, as discussed
below. Therefore, we will model tidal streams in frequency--angle
space. Because the leading and trailing arm are well-separated in
frequency space, we model each arm individually.

A generative model of a tidal stream in frequency--angle space
requires three ingredients: (a) a model for the distribution of times
at which stars are stripped from the progenitor, (b) a prescription
for the distribution of frequency offsets from the progenitor at every
given stripping time, and (c) a description of the angle offsets for
any given frequency offset and stripping time. Having specified these
ingredients, we can then generate tidal debris at any given stripping
time and evolve it forward using the simple linear dynamics in
frequency--angle space discussed in
\sectionname~\ref{sec:dynamics}. The initial angle offsets are small
enough that they are likely unobservable even with futuristic data
(especially in the direction of the frequency offset, where the
initial angle offset is quickly overwhelmed by the subsequent
dynamical evolution; for the simulated stream used here this happens
after approximately $20\Myr$). For that reason, I will assume that the
initial angle offsets are independent of both stripping time and
frequency offset. In what follows, I will model them using a simple
isotropic Gaussian distribution with a dispersion $\sigma_\theta$.

\begin{figure*}[t!]
  \includegraphics[width=0.32\textwidth,clip=]{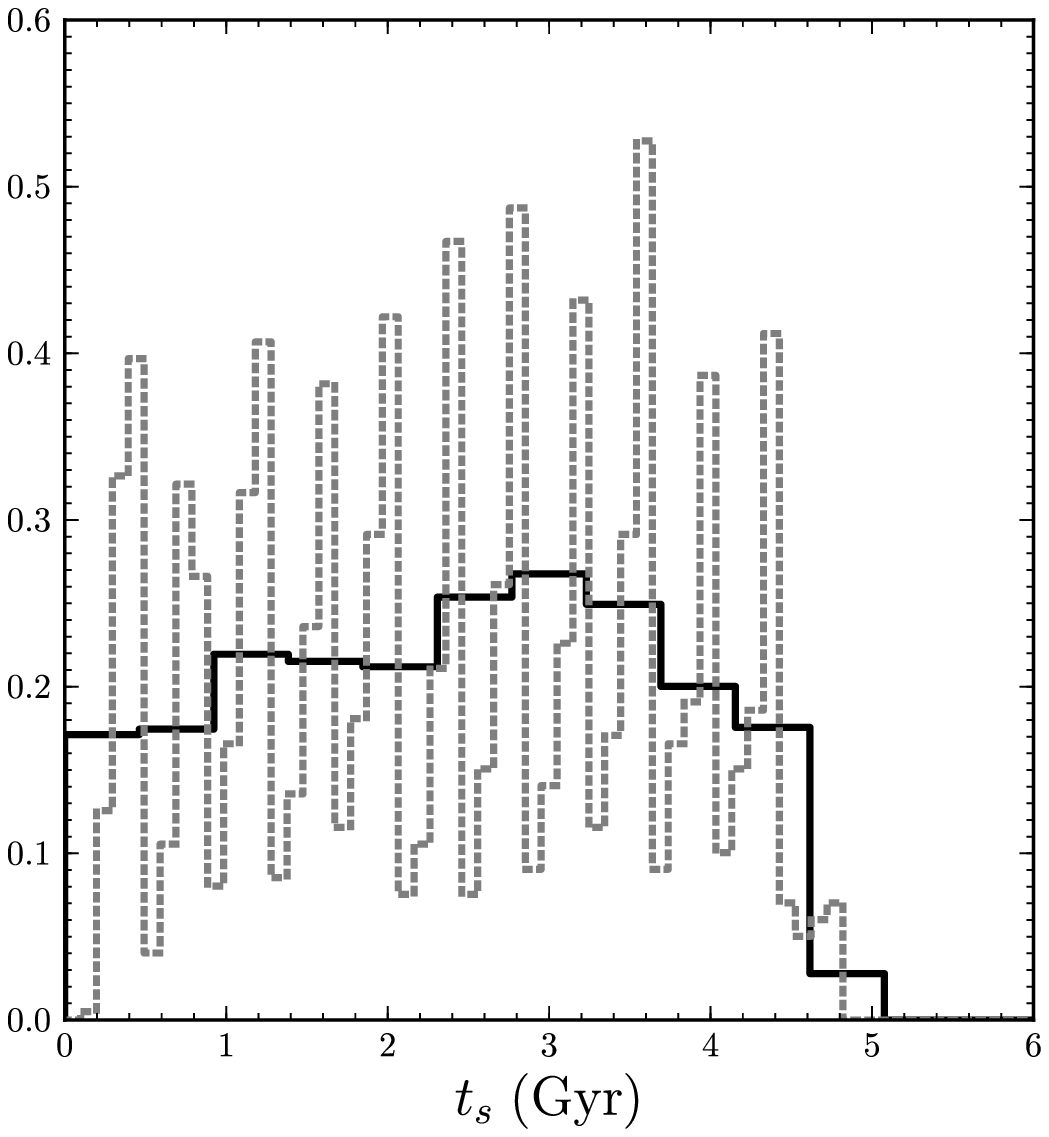}
  \includegraphics[width=0.32\textwidth,clip=]{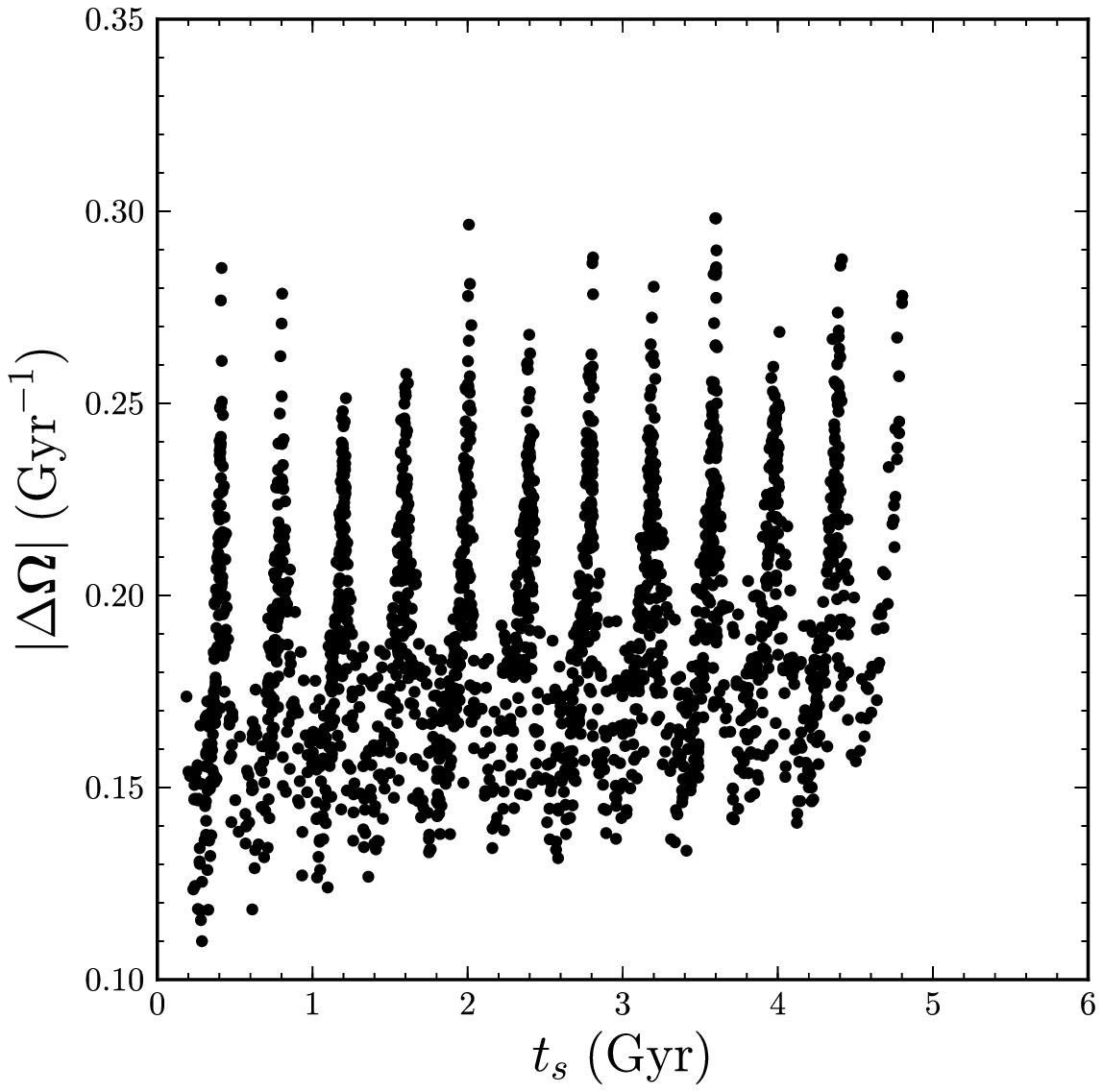}
  \includegraphics[width=0.32\textwidth,clip=]{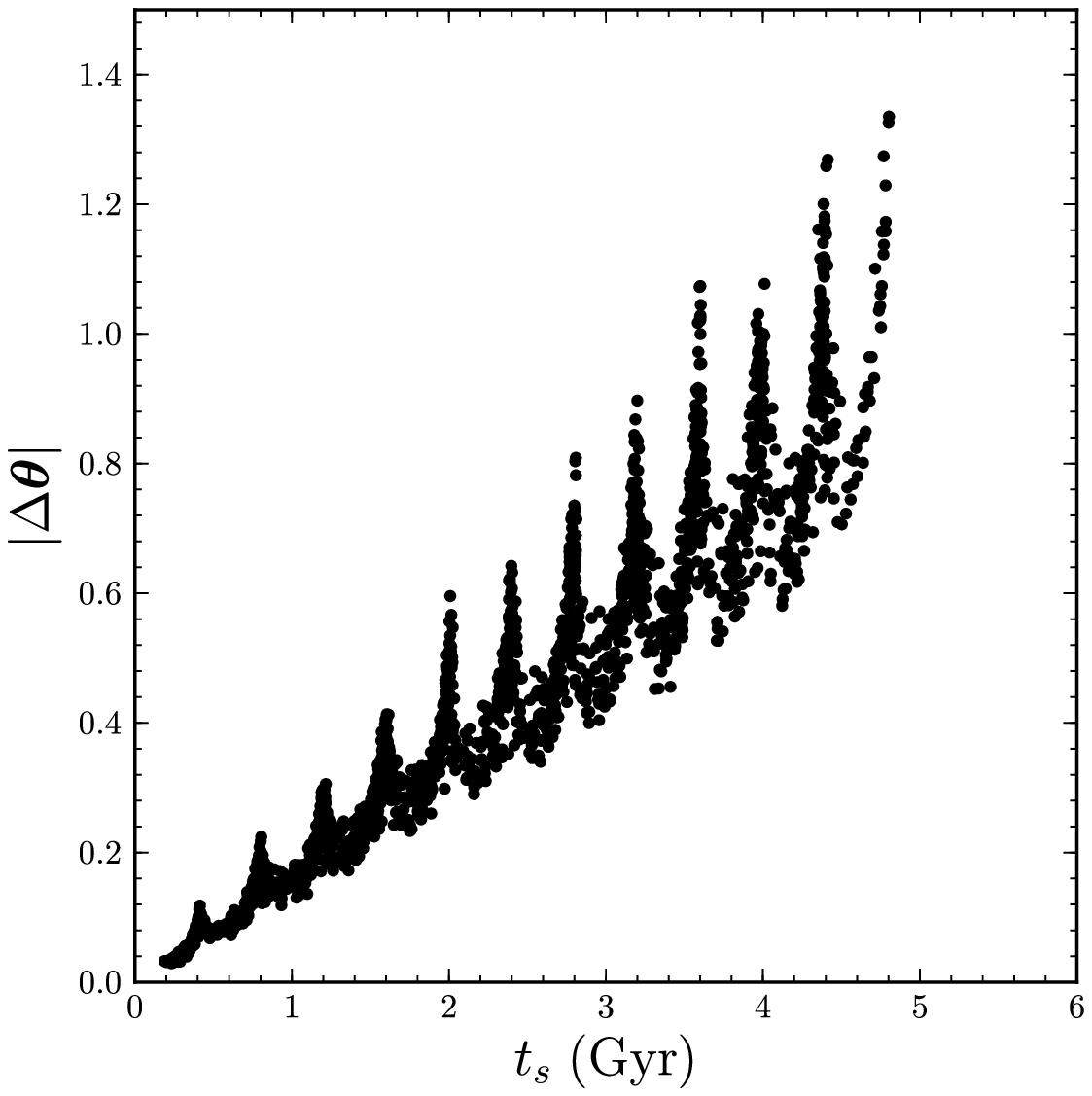}
  \caption{Properties of the simulated tidal stream of
    \sectionname~\ref{sec:sim} as a function of time. Stripping times
    are calculated by linear regression of the angle difference
    between a stream particle and the progenitor versus the equivalent
    frequency difference. The distribution of stripping times is shown
    in the left panel.  While the stripping process happens in bursts
    with a period of roughly $400\Myr$ (the radial period of the
    progenitor), it can be approximated as constant (more
    coarsely-binned solid histogram). The magnitude of the frequency
    offset between stream particles and the progenitor is shown in the
    middle panel. While particles stripped at pericenter have larger
    frequency offsets than those between pericenter passages, there is
    almost no secular trend and the typical frequency offset (averaged
    over a radial period) can be considered constant with time. The
    right panel shows the magnitude of the angle offset between stream
    particles and the progenitor as a function of stripping
    time. There is a spread in the angle offset at all stripping times
    because of the dispersion in frequency offsets, but typically
    stars that have been stripped earlier reach larger angle
    offsets. That is, self-sorting happens only to a small
    degree.}\label{fig:gd1_times}
\end{figure*}

Ingredients (a) and (b) require more careful modeling. The right panel
of \figurename~\ref{fig:gd1_times} shows the distribution of times at
which stream particles are stripped from the progenitor. These
stripping times are calculated from the final snapshot by linear
regression of a particle's angle offset $\Delta \veca$ from the
progenitor, versus the frequency offset $\Delta \veco$ as 
\begin{equation}
  \ts = \frac{\Delta \veco \cdot \Delta \veca}{|\Delta \veco|^2}\,.
\end{equation}
The dotted line shows a finely-binned histogram of these stripping
times, which indicates that stripping happens in bursts with a period
of about $400\Myr$, the radial period of the progenitor's orbit. Thus,
as expected, stripping happens primarily at pericenter passages, with
a smaller number of particles lost between pericenter
passages. Realistic models of a stream, especially those that require
good models of the surface-density structure of the stream as a
function of stream angle, need to model this non-uniform stripping
process. In what follows, we will use a uniform distribution of
stripping times. This is appropriate for modeling the large-scale
structure of streams, such as what is used when constraining the
gravitational potential.

The middle panel of \figurename~\ref{fig:gd1_times} shows the
frequency offsets between stream particles and the progenitor as a
function of stripping time, corresponding to ingredient (b) above. The
same burstiness that is apparent in the distribution of stripping
times shows up here. Particles stripped at pericenter typically have
larger frequency offsets than those lost at larger radii. This can be
described as saying that particles are \emph{removed} at pericenter,
but only \emph{peeled off} between pericenter passages. Again,
realistic models of tidal streams need to take this detailed structure
into account.

The right panel of \figurename~\ref{fig:gd1_times} shows the angle
offset (at the end of the simulation) versus the stripping time. There
is a spread in the angle offset reached for any given stripping time,
but it is nevertheless typically the case that particles that have
been stripped earlier are found at larger angle offsets
today. Self-sorting---the tendency of particles stripped later with
larger frequency offsets to overtake particles stripped earlier at
smaller frequency differences and thus to erase correlations between
particle position along the stream and stripping time and sort the
stream in frequency offset---happens to a small extent but the narrow
distribution of frequency offsets limits the ability of particles
removed later to reach large distances from the progenitor.

We can model streams in frequency space because the mapping $\vecj
\rightarrow \veco$ is of full rank for axisymmetric and triaxial
potentials. However, for spherical potentials this mapping only has
rank two, because $\Omega_\phi = \mathrm{sgn}(L_Z)\,\Omega_Z$, such
that the determinant $|\partial \veco / \partial \vecj|$ is zero. For
spherical potentials, tidal streams only spread in a five-dimensional
subspace and when modeling streams in spherical potentials we need to
keep the direction perpendicular to this subspace constant.

\subsection{The fiducial stream model}\label{sec:fidmodel}

The discussion of the simulated stream's properties as a function of
stripping time in the previous section indicates that we can build
simple models of the structure of tidal streams when considering them
in frequency--angle space. In this section, I propose a very simple
fiducial model that I will use in the remainder of this paper. This
model combines simple analytic forms for ease of computation that are
nevertheless realistic enough to provide adequate models for tidal
streams in many applications.

First, we approximate the bursty, non-uniform distribution of
stripping times in \figurename~\ref{fig:gd1_times} with a uniform
distribution
\begin{align}\label{eq:pt}
  \begin{split}
    p(\ts) & \propto 1\,,\qquad 0 < \ts < t_d\\
    & = 0\,,\qquad \mathrm{otherwise}\,.
  \end{split}
\end{align}
The thick, solid histogram in the right panel of
\figurename~\ref{fig:gd1_times} shows a coarser-binned histogram of
stripping times, indicating that the uniform distribution is a good
approximation over intervals longer than a radial period ($\approx
400\Myr$). The \emph{disruption time} $t_d$ is a free parameter of the
model that needs to be determined for each stream. For the model of
the simulated stream used here I set $t_d = 4.5\Gyr$. This
distribution of stripping times is different from that of
\citet{Johnston98a}, who assumed that stripping happens exactly at
pericenter and is therefore a sum of delta functions located at each
pericenter passage. We do not consider such a distribution here
further, but it should be straightforward to repeat the analytical
calculations in \sectionname~\ref{sec:track} and \ref{sec:pdf} below
for this alternative distribution.

Similarly, we model the distribution of frequency offsets between
stream particles and the progenitor to be independent of stripping
time, even though the middle panel of \figurename~\ref{fig:gd1_times}
clearly shows that particles are stripped at larger frequency
differences at pericentric passages than between them. The fact that
there is only a small long-term trend in the typical frequency offset
means that this is a good model on time-scales larger than the radial
period. While this means that the small-scale structure of the stream
will not be perfectly represented by the model, the large-scale
structure of the stream will be fine. For using tidal streams to
constrain the gravitational potential, large arcs are most important,
such that this simple model will be adequate for such inferences.

The distribution of frequencies is three dimensional. Because we model
the leading and the trailing arm separately, a single-peaked
distribution suffices to describe the frequency distribution. A
general description in terms of a Gaussian distribution would have
nine free parameters. However, we can model it with fewer free
parameters as follows. First, following \citet{Eyre11a}, we
approximate the \emph{action} distribution as a Gaussian with standard
deviations given by the approximate spread in the actions $(\delta
J_R,\delta L_Z,\delta J_Z)$ in the cluster
\begin{align}
  \sigma_{J_R} & \approx \delta \left(\frac{1}{2\,\pi}\oint\dd r\,p_r\right) \approx \frac{1}{\pi}\,\sigv\,\left(r_{\mathrm{apo}}-r_{\mathrm{peri}}\right)\,,\\
  \sigma_{L_Z} & \approx  \delta L_Z \approx \sigv\,r_{\mathrm{peri}}\,,\\
  \sigma_{J_Z} & \approx \delta\left(\frac{1}{\pi}\oint\dd z\,p_z \right)\approx \frac{2}{\pi}\,\sigv\,Z_{\mathrm{max}}\,,
\end{align}
where $\sigv$ is the velocity dispersion of the progenitor,
$r_{\mathrm{apo}}$ and $r_{\mathrm{peri}}$ are the apo- and pericenter
radii of the progenitor orbit, respectively, and $Z_{\mathrm{max}}$ is
the maximum $Z$ reached on this orbit. We assume that all of the
correlations between the actions are zero (see
\figurename~\ref{fig:gd1_jt}). Then we propagate this Gaussian
variance to frequency space using the Hessian $(\partial \veco /
\partial \vecj)\large|_{\vecj^p}$\footnote{The Hessian $(\partial
  \veco / \partial \vecj)\large|_{\vecj^p}$ can be computed using the
  method in \appendixname~\ref{sec:aa} by computing $\partial
  (\vecj,\veca) / \partial (\vecx,\vecv)$ and $\partial (\veco,\veca)
  / \partial (\vecx,\vecv)$ and forming $\partial \veco / \partial
  \vecj = \left(\partial (\vecj,\veca) / \partial
  (\vecx,\vecv)\right)^{-1}\,\left(\partial (\veco,\veca) / \partial
  (\vecx,\vecv)\right)$.}, which gives the variance matrix
$V_\veco$. The principal eigenvector of this matrix is the model
direction along which the stream spreads. While the model I propose
here is a simple one, it works well without having been tweaked for
the simulation under scrutiny. The angle between the principal
eigenvector of $V_\veco$ and the progenitor's frequency vector is
$0.50^\circ$ which is close to the value measured from the simulation
using the mean frequency-offset vector is $0.18^\circ$ (see
\figurename~\ref{fig:gd1_jt}). For a model with an isotropic action
distribution, the misalignment would be $1.28^\circ$, much larger than
the measured value.

\begin{figure}[t!]
  \includegraphics[width=0.44\textwidth,clip=]{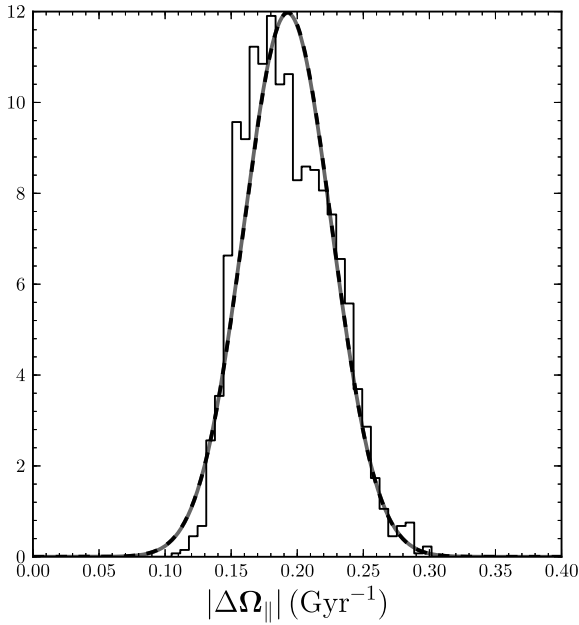}
  \caption{Distribution of the magnitude of frequency offsets between
    stream particles and the progenitor along the direction of the
    mean frequency offset. The dashed line is a Gaussian fit to this
    distribution, while the solid line is a fit of our preferred form
    for this distribution, given in \equationname~(\ref{eq:do}) (the
    dashed and solid lines almost entirely overlap). Both forms
    provide adequate fits to the distribution. }\label{fig:gd1_dohist}
\end{figure}

I then further constrain the mean frequency offset to lie along the
principal eigenvector of $V_\veco$. Therefore, this mean frequency
offset can be described in terms of a parameter $\mu_\Omega = \Delta
\Omega^m / \sigma_{\Omega,1}$, where $\sigma_{\Omega,1}$ is the square
root of the largest eigenvalue of $V_\veco$ and $\Delta \Omega^m$ is
the one-dimensional mean frequency offset. The sign of $\mu_\Omega$
sets whether we are modeling a leading or a trailing arm. We can gain
insight on good values for $\mu_\Omega$ from the
simulation. \figurename~\ref{fig:gd1_dohist} shows the distribution of
frequency offsets in the direction of the median frequency offset in
the stream. Two fits to this distribution are displayed in
\figurename~\ref{fig:gd1_dohist}. The dashed line is a Gaussian fit to
this distribution, while the solid line is a fit of the form
\begin{equation}\label{eq:do}
  p(|\Delta \veco|_\parallel) \propto |\Delta \veco|_\parallel\,\mathcal{N}\left(|\Delta \veco|_\parallel|\Delta \Omega^m,\sigma_{\Omega,1}^2\right)\,,
\end{equation}
where $\mathcal{N}(x|m,v)$ represents a Gaussian distribution for $x$
with mean $m$ and variance $v$. The fit is equally good (because the
dispersion is much smaller than the mean) and I choose the second form
because it simplifies some of the stream-track calculations below. The
best-fit $(\Delta \Omega^m,\sigma_{\Omega,1})$ is
$(0.19,0.033)\Gyr\inv$ for both fits. Based on this information, a
model with $\sigv = 0.365\kms$ and $\mu_\Omega = 6$ provides a good
fit to the simulation data. The distribution of frequency offsets in
the two-dimensional space perpendicular to the principal eigenvector
of $V_\veco$ is modeled as a zero-mean Gaussian with variance matrix
given by the projection of $V_\veco$ onto this space.

Because we expect the typical $\Delta\veco$ to scale with the velocity
dispersion of the progenitor, I conjecture that a constant
$\mu_\Omega$ can be used for modeling tidal streams of any (small-ish)
velocity dispersion. Further modeling of simulated streams, however,
is necessary to check this and to find the optimal value of
$\mu_\Omega$. In particular, if the progenitor has internal rotation
the action distribution of the debris and $\mu_\Omega$ may be
significantly different. Fixing $\mu_\Omega$ leaves a single free
parameter \sigv\ for modeling the frequency offset distribution of a
tidal stream. In modeling observed data, this parameter needs to be
fit and we expect it to be proportional to the velocity dispersion of
the progenitor, although this proportionality needs to be checked more
carefully. The discussion in \sectionname~5 and Figure 3 in
\citet{Sanders13a} demonstrates that the size of the frequency
distribution scales as mass$^{1/3}$ over more than five orders of
magnitude in mass, or approximately as the velocity dispersion through
application of the virial theorem and the expression for the tidal
radius as a function of the progenitor mass. Similarly,
\citet{Johnston98a} successfully modeled streams by assuming that the
scale of the energy distribution in the stream is proportional to
mass$^{1/3}$. The initial angle distribution, modeled here as an
isotropic Gaussian, has a characteristic spread $\sigma_{\theta}$
which I set here to $\sigma_\theta=0.003$ ($=\sigv/[122\kms]$) for the
model $\sigv=0.365\kms$; see above) based on a comparison with the
simulation data, and we likewise expect this spread to scale with the
velocity dispersion of the progenitor.

\begin{figure*}[t!]
\begin{center}
  \includegraphics[width=0.32\textwidth,clip=]{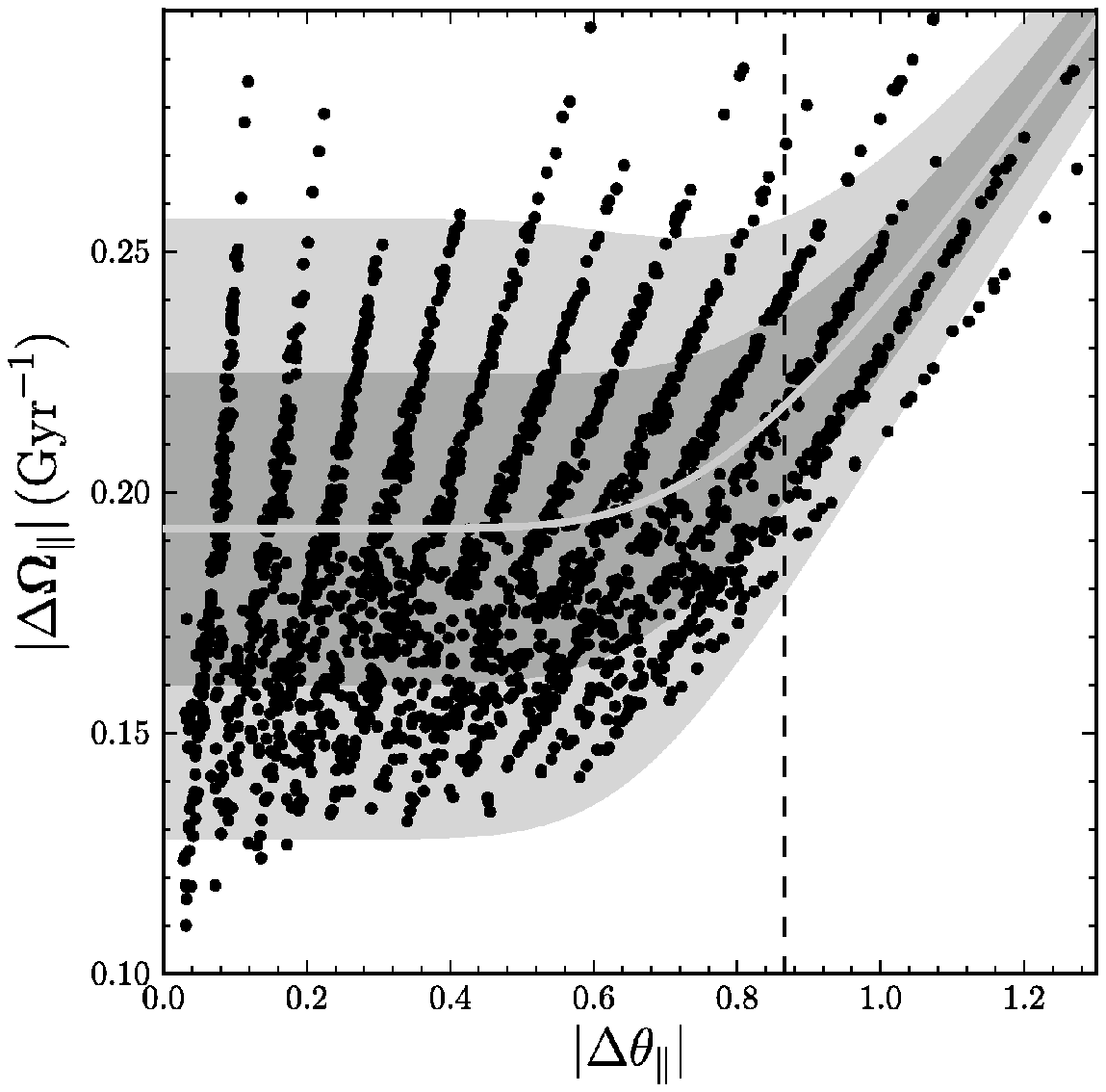}
  \includegraphics[width=0.32\textwidth,clip=]{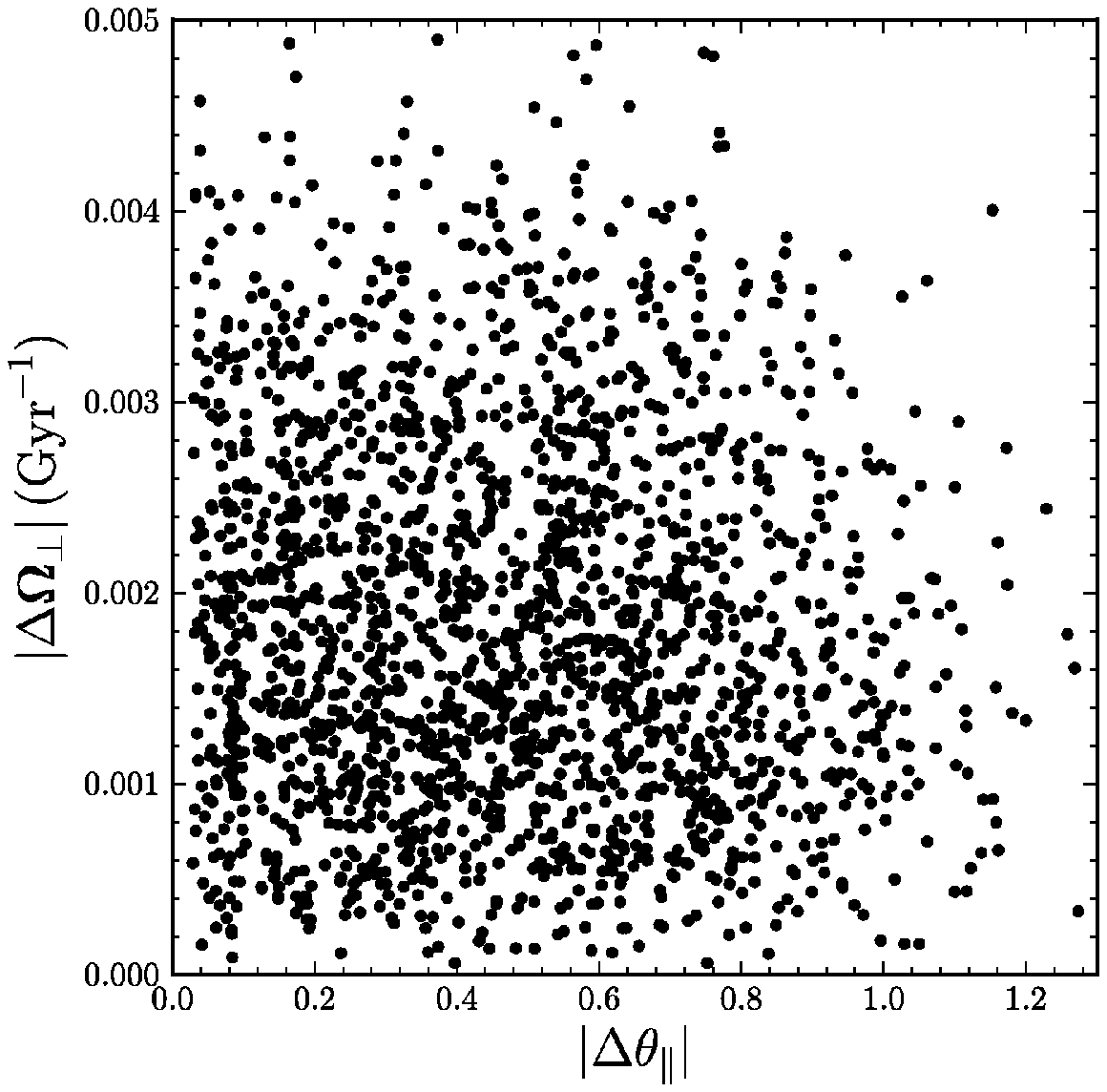}\\
  \includegraphics[width=0.32\textwidth,clip=]{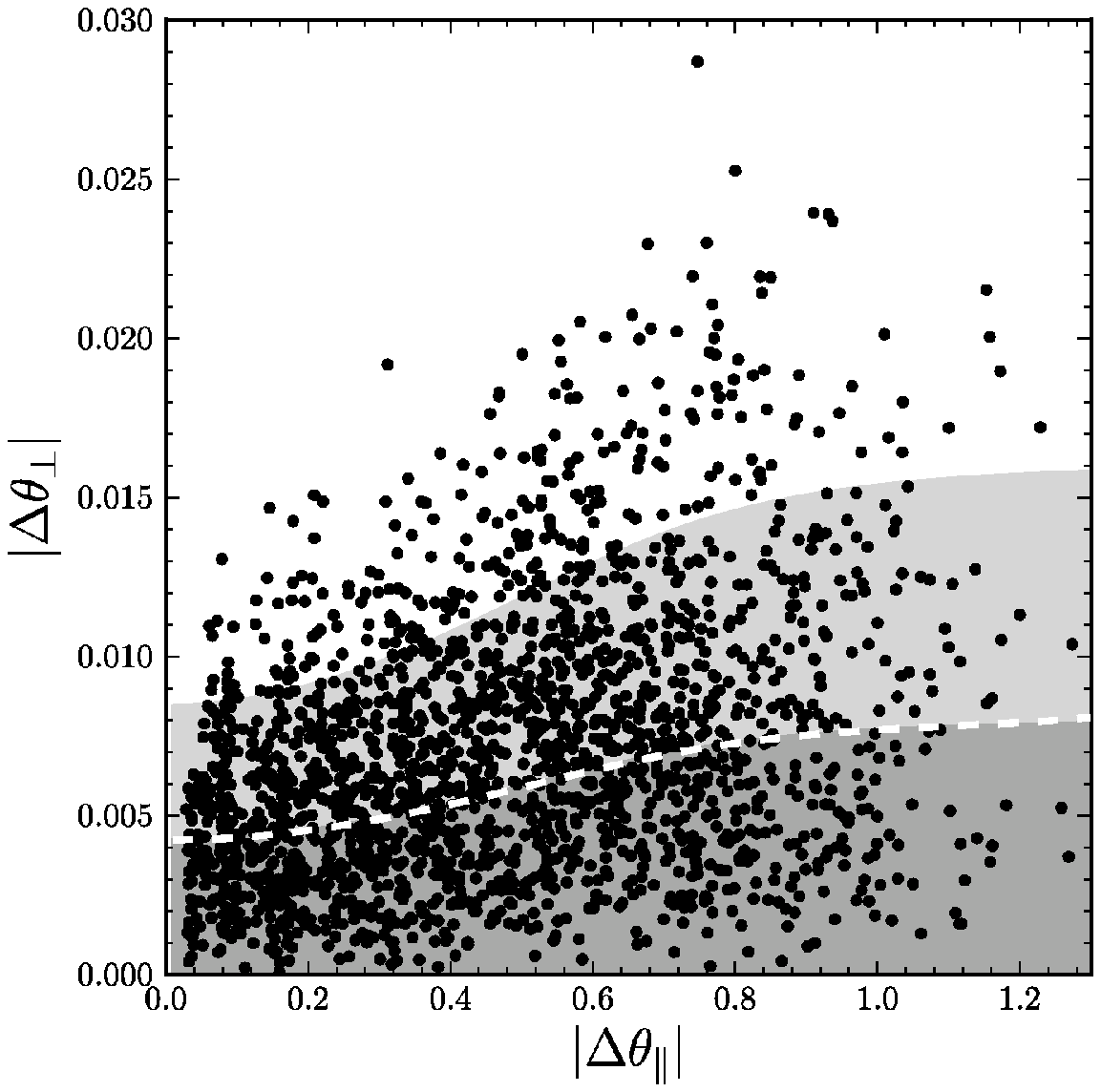}
  \includegraphics[width=0.32\textwidth,clip=]{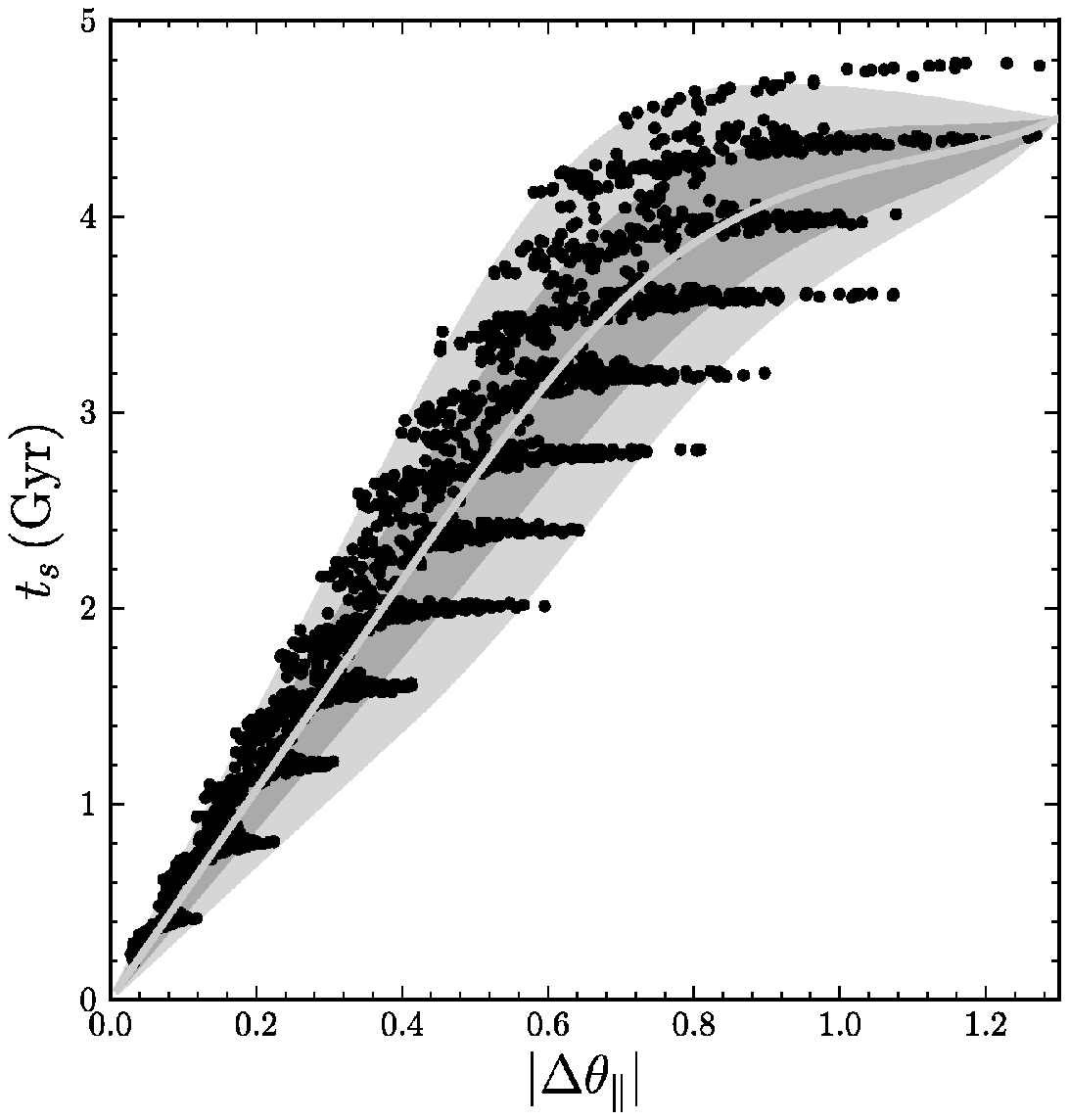}
  \caption{Stream properties as a function of angle along the
    stream. The top, left panel shows the magnitude of frequency
    offsets along the stream. The gray line shows the mean predicted
    frequency offset calculated using the framework of
    \sectionname\sectionname~\ref{sec:modeloa} and \ref{sec:track},
    while the grayscale bands show the $68\,\%$ and $95\,\%$ limits of
    the predicted distribution. The dashed line shows the angle $\apar
    = \Delta \Omega^m\,t_d$, angles beyond which can only be reached
    by having a larger $\Delta \opar$ than average. The top, right
    panel shows the magnitude of frequency offsets perpendicular to
    those in the left panel. The predicted distribution is constant as
    a function of angle and therefore not shown. The bottom, left
    panel shows the angle offset perpendicular to the stream and the
    grayscale bands show $68\,\%$ and $95\,\%$ limits of the predicted
    distribution of angle offsets; the white dashed line shows the
    $68\,\%$ limit calculated using a simplified method. The bottom,
    right panel shows the stripping time. The gray line and grayscale
    bands again show the mean and $68\,\%$ and $95\,\%$ confidence
    limits of the predicted distribution. While the ``burstiness''
    related to the energetic stripping at pericenter that is apparent
    in the simulated data's distributions is (by design) absent in the
    model, the overall scale and the mean trends with angle are well
    represented by the model.}\label{fig:gd1_apar}
\end{center}
\end{figure*}

The fiducial model for a tidal stream in frequency--angle coordinates
proposed here is therefore described by essentially two free
parameters in addition to the progenitor's phase-space position: the
progenitor's velocity dispersion \sigv\ and the disruption time
$t_d$. As shown below, these two parameters are the most important in
determining the track of the stream. More sophisticated models can
leave $\mu_\Omega$ and $\sigma_{\theta}$ as free parameters as
well. In what follows, I will fix these to the values given in the
previous paragraphs. It must be stressed, however, that all of these
parameters here have been fixed ``by-eye'', and better fits might be
possible using more quantitative fit procedures.

\section{The track of a tidal stream}\label{sec:track}

In this Section, I discuss how to calculate the track of a model tidal
stream in the generative model described in
\sectionname~\ref{sec:generative}. The track consists of the mean
location of the stream as a function of angle along the stream. I also
describe how to estimate the dispersion of the stream along this
track. In \sectionname~\ref{sec:trackaa}, I explain how to calculate
the track of the stream in frequency--angle coordinates as well as how
to estimate the spread around this track. In
\sectionname~\ref{sec:trackxv} I discuss how to efficiently project
the stream track and dispersion into position--velocity coordinates
$(\vecx,\vecv)$ or observable coordinates $(l,b,D,\vlos,\pmll,\pmbb)$.

\begin{figure*}[t!p]
\begin{center}
  \includegraphics[width=0.8\textwidth,clip=]{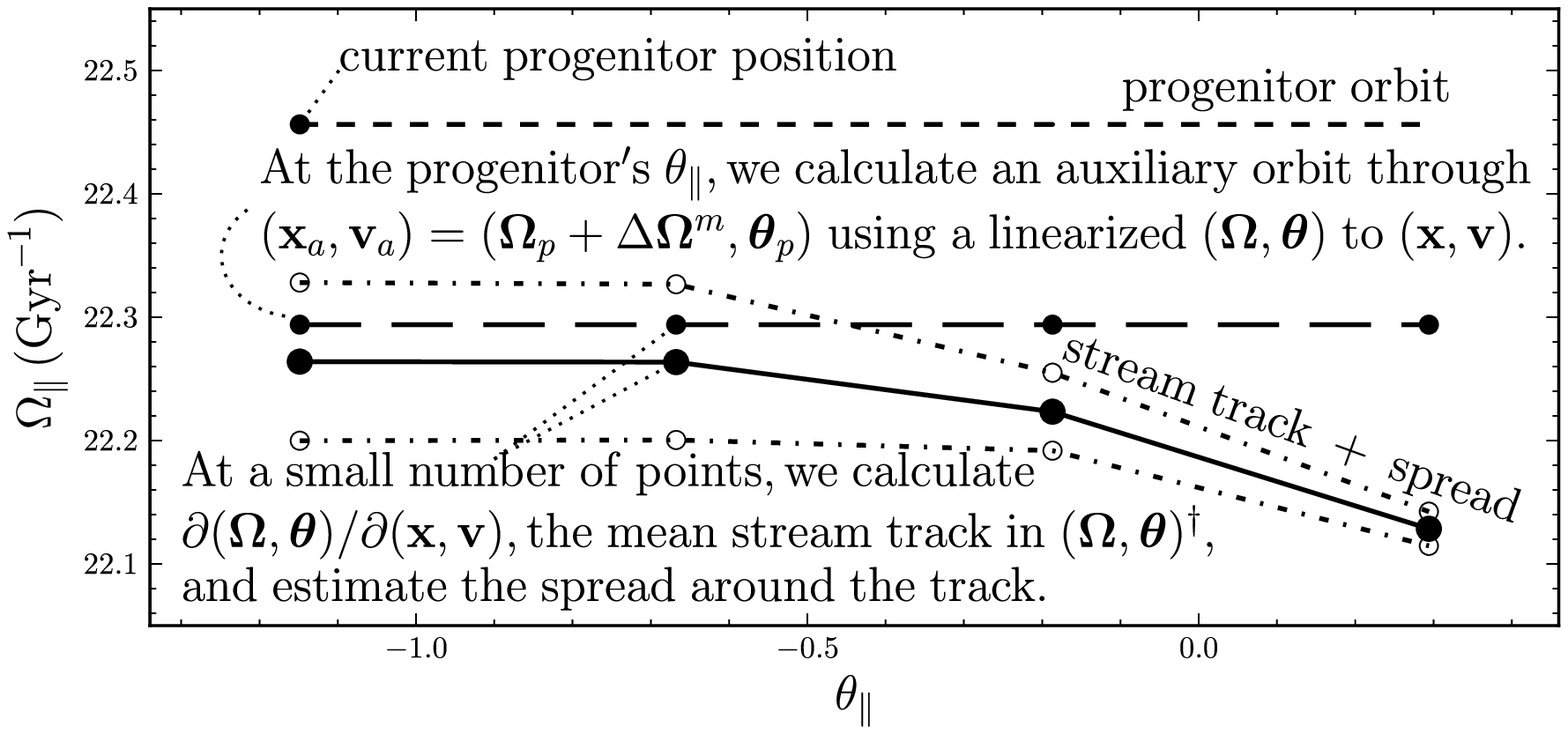}\\
  \includegraphics[width=0.8\textwidth,clip=]{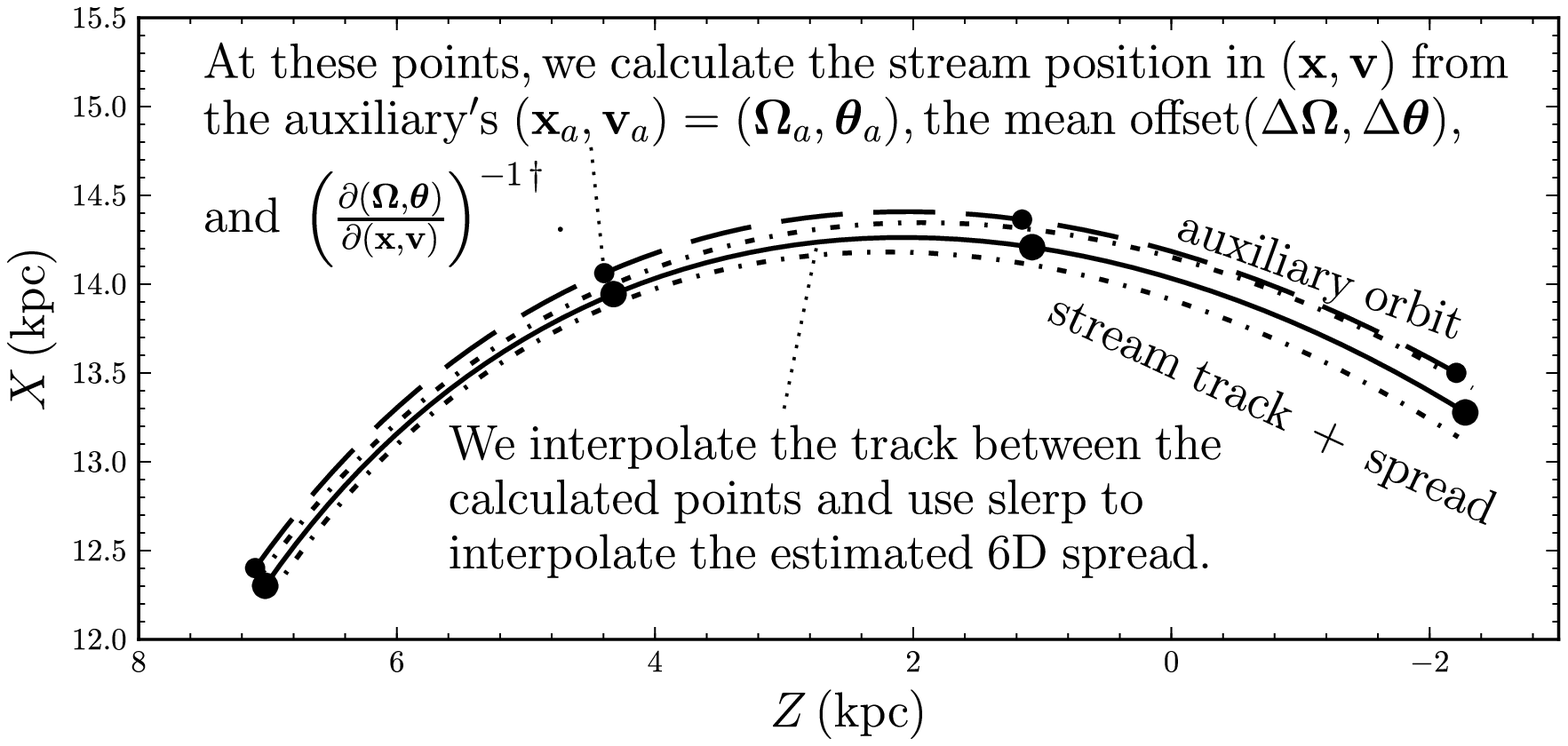}\\
  \includegraphics[width=0.8\textwidth,clip=]{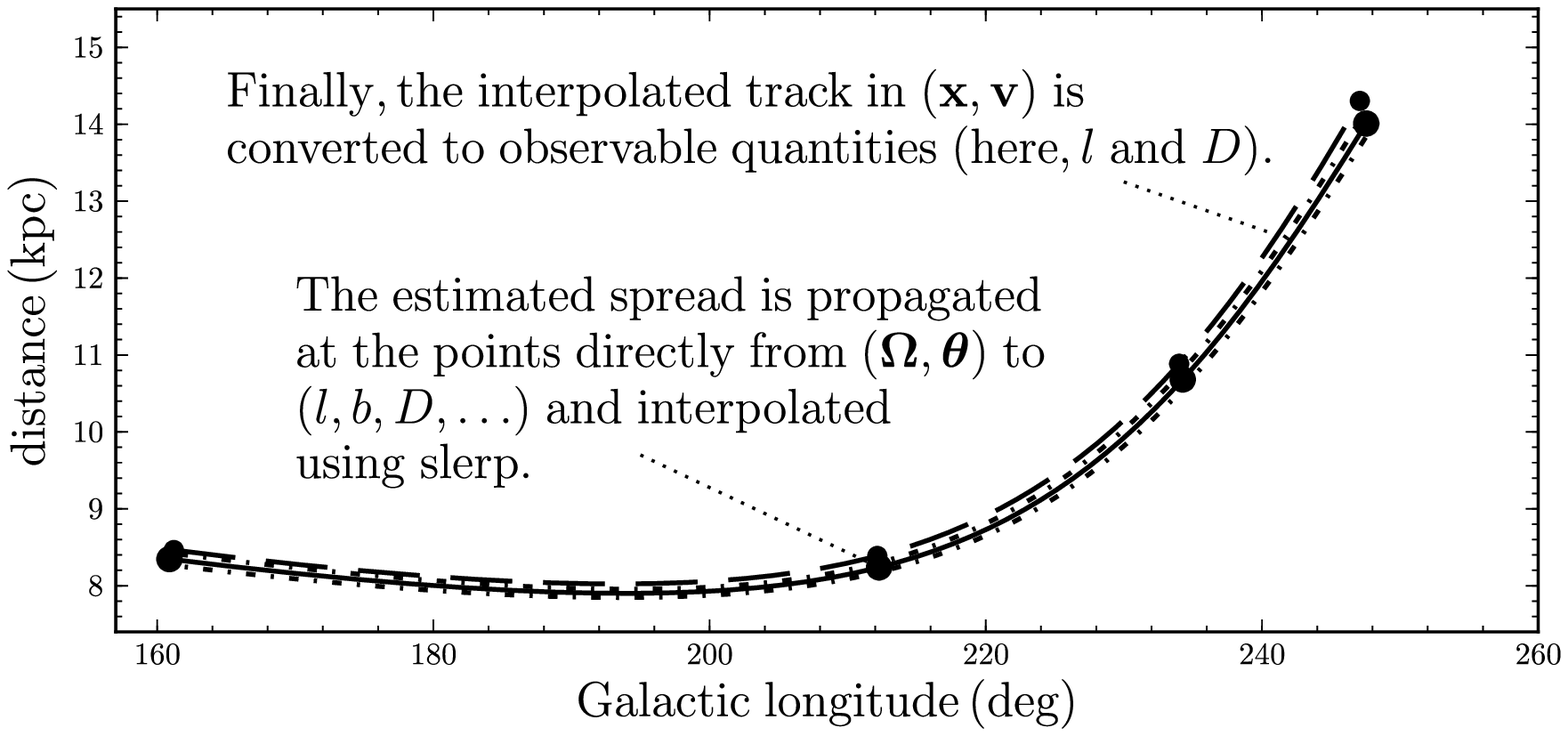}
  \caption{Illustration of the determination of the average stream
    track described in \sectionname~\ref{sec:track}. The small black
    dots in each panel are the points along the auxiliary orbit at
    which the Jacobian $\partial (\veco,\veca) / \partial
    (\vecx,\vecv)$ is calculated. The large black dots are the
    corresponding points on the stream where the track is
    calculated. The short-dashed line in each panel is the
    progenitor's orbit, the long-dashed line is the auxiliary orbit,
    and the solid line and dash-dotted lines are the mean track and
    $2\sigma$ dispersion. The daggers ($\dagger$) indicates the step in
    this procedure where the calculation of the mean stream track in
    $(\vecx,\vecv)$ can be iterated (see text). The offset between the
    auxiliary orbit and the stream track is exaggerated for display
    purposes in all panels; typically the auxiliary orbit is very
    close to the stream track.}\label{fig:track}
\end{center}
\end{figure*}

\subsection{In frequency--angle coordinates}\label{sec:trackaa}

In the generative model described above, the direction of the angle
along the stream lies along the principal eigenvector of $V_\veco$,
the variance matrix of the model $\Delta \veco$ distribution. We
denote the angle in this direction as $\apar$ and the frequency in
this direction $\opar$; the two-dimensional angle- and frequency-space
perpendicular to this direction is denoted as $\aperp$ and $\operp$.

First, we compute the distribution of $\Delta \opar$, the frequency
offset between stream members and the progenitor given the angle
offset $\Delta \apar$ as follows (considering $\Delta \opar$ to be
positive)
\begin{equation}\label{eq:poparapargeneral}
\begin{split}
  p(\Delta \opar|\Delta \apar) & = \int \dd \ts\,p(\Delta \opar,\ts|\Delta \apar)\, \\
  & \propto \int \dd \ts \,p(\Delta \apar | \Delta \opar,\ts)\,p(\Delta \opar,\ts)\,,\\
  & \approx \int \dd \ts \,\delta(\Delta \apar - \Delta \opar\,\ts)\,p(\Delta \opar,\ts)\,,\\
  & = \frac{1}{\Delta \opar}\,p(\Delta \opar,\frac{\Delta \apar}{\Delta \opar})\,,
\end{split}
\end{equation}
where in the penultimate step I have approximated the initial angle
distribution as a delta function (as the initial angle offset is small
with respect to the final offset). This is a fully general
expression. For the fiducial stream model, this can be simplified to
\begin{equation}\label{eq:poparapar}
\begin{split}
  p(\Delta \opar|\Delta \apar) \propto &\ \frac{1}{\Delta \opar}\,p(\Delta \opar)\,,\qquad 0 < \frac{\Delta \apar}{\Delta \opar} < t_d\,.\\
  & 0\,,\qquad \mathrm{otherwise}\,.
\end{split}
\end{equation}
\Eqnname~(\ref{eq:poparapar}) shows why I chose to model $p(\Delta
\opar)$ as a Gaussian multiplied with $|\Delta \opar|$. $p(\Delta
\opar|\Delta \apar)$ is then a Gaussian with mean $\Delta \Omega^m$
and variance $\sigma^2_{\Omega,1}$, for $\Delta \opar > \Delta \apar /
t_d$, and zero otherwise. The mean and variance of such a Gaussian is
straightforward to calculate in terms of error functions. The top left
panel of \figurename~\ref{fig:gd1_apar} shows the distribution of
$|\Delta \opar|$ versus $|\Delta \apar|$ (absolute value in order to
show the trailing and leading stream together) for the simulated
stream, as well as the mean and dispersion calculated from the
fiducial model. The dashed line shows $\Delta \apar = \Delta \Omega^m
\,t_d$, approximately the angle where the finite disruption time
starts to influence the mean $\Delta \opar$ of the stream. The finite
disruption time, $\approx 4.5\Gyr$ in this case, of a tidal stream
means that for stream members to have reached very large angle
differences with respect to the progenitor, they must have been
removed at large frequency differences. Around $\Delta \Omega^m
\,t_d$, average stream stars, \ie, those removed with the average
frequency offset, do not have a large enough offset to reach large
angle offsets. Therefore, even though the average frequency offset
does not change much over time (see \figurename~\ref{fig:gd1_times}),
the average frequency offset, and therefore the average orbit, does
change with distance from the progenitor.

In the fiducial model, the average $\Delta \operp$, and $\Delta
\aperp$ as a function of $\Delta \apar$ are zero. Therefore, the
average stream track as a function of $\Delta \apar$ is entirely
specified by the mean $\Delta \opar(\Delta \apar)$ in the parallel
direction and zero offsets in the perpendicular directions. This track
can be rotated into the $(R,\phi,Z)$ coordinate system using the
eigenvectors of $V_\veco$.

To estimate the spread around the stream track, we need to calculate
the distributions $p(\Delta \operp|\Delta \apar)$ and $p(\Delta
\aperp|\Delta \apar)$. The former is just the zero mean Gaussian with
variance given by the projection of $V_\veco$ onto the direction
perpendicular to the stream. The distribution $p(\Delta \aperp|\Delta
\apar)$, however, does depend on $\Delta \apar$, because $\Delta
\aperp = \Delta \aperp^{\mathrm{init}} + \Delta \operp \ts$ and
the distribution of $\ts$ depends on $\Delta \apar$. Therefore,
we first calculate $p(\ts | \Delta \apar)$.

We can calculate $p(\ts | \Delta \apar)$ as follows
\begin{equation}\label{eq:pdtapar}
\begin{split}
  p(\ts | \Delta \apar) & = \int \dd \Delta \opar\,p(\ts,\Delta \opar|\Delta \apar)\,,\\
  & = \int \dd \Delta \opar\,p(\ts|\Delta \opar,\Delta \apar)\,p(\Delta \opar|\Delta \apar)\,,\\
  & \approx p\left(\Delta \opar = \frac{\Delta \apar}{\ts}\right)\,\left|\frac{\Delta \apar}{(\ts)^2}\right|\,,
\end{split}
\end{equation}
where in the last step I have again approximated the initial
distribution of parallel angle offsets as a delta function. The first
factor in this equation is given by the expression given in
\equationname~(\ref{eq:poparapargeneral}) (in general) or in
\equationname~(\ref{eq:poparapar}) for the fiducial model. The lower
right panel of \figurename~\ref{fig:gd1_apar} shows the distribution
of stripping times for members of the simulated stream as well as the
mean and dispersion calculated from
\equationname~(\ref{eq:pdtapar}). Similar to the distribution of
$\Delta \opar$ as a function of $\Delta \apar$, the distribution of
$\ts$ has a kink at the angle offset that can only be reached by stars
that have to have been stripped at large frequency offsets (that is,
larger than average).

We can now also calculate $p(\Delta \aperp|\Delta \apar)$:
\begin{equation}
\begin{split}
  p(\Delta \aperpii|\Delta \apar) & = \int \dd \ts \,\dd \Delta
  \operpii \,p(\Delta \aperpii|\Delta \operpii,\ts)\\
  & \qquad \times p(\Delta \operpii|\ts,\Delta \apar)\,p(\ts|\Delta \apar)\,,
\end{split}
\end{equation}
where $ii=2,3$ for the first and second perpendicular direction. For
the fiducial model, this simplifies to
\begin{equation}\label{eq:paperpapar}
\begin{split}
  p(\Delta \aperpii|\Delta \apar) & = \int \dd \ts \,\mathcal{N}\left(\Delta \aperpii|0,\sigma_{\theta}^2+(\ts\,\sigma_{\Omega,ii})^2\right)\qquad\\
& \qquad \times p(\ts|\Delta \apar)\,,
\end{split}
\end{equation}
where $\sigma_{\Omega,ii}$ is the model frequency dispersion in the
perpendicular direction (corresponding to the middle and smallest
eigenvalue of $V_\veco$). I have \emph{not} approximated the initial
angle-offset distribution as a delta function here, because its width
is a significant fraction of that of the final
perpendicular-angle-offset distribution. In the fiducial model, this
distribution is the only one of those considered in this section that
requires an explicit integration. However, we can approximate the
distribution by using the mean $\ts(\Delta \apar)$ rather than
integrating over $\ts$. The lower left panel of
\figurename~\ref{fig:gd1_apar} shows the distribution of perpendicular
angle offsets and the grayscale bands show the dispersion calculated
using \equationname~(\ref{eq:paperpapar}); the white dashed line shows
the approximate estimate obtained using the mean stripping time rather
than integrating over it. This simple estimate agrees well with the
exact calculation.

\begin{figure*}[t!]
\begin{center}
  \includegraphics[width=0.8\textwidth,clip=]{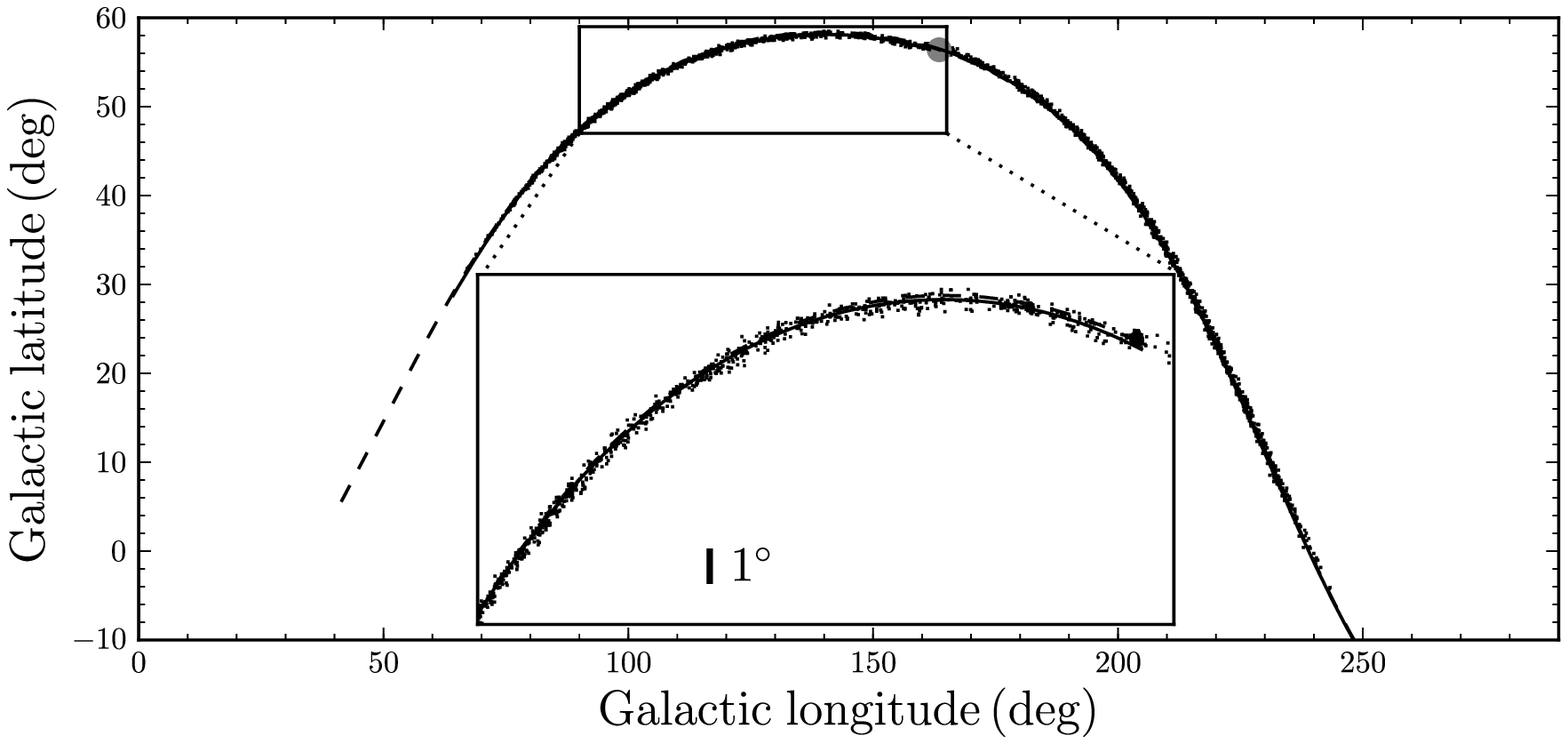}
  \includegraphics[width=0.8\textwidth,clip=]{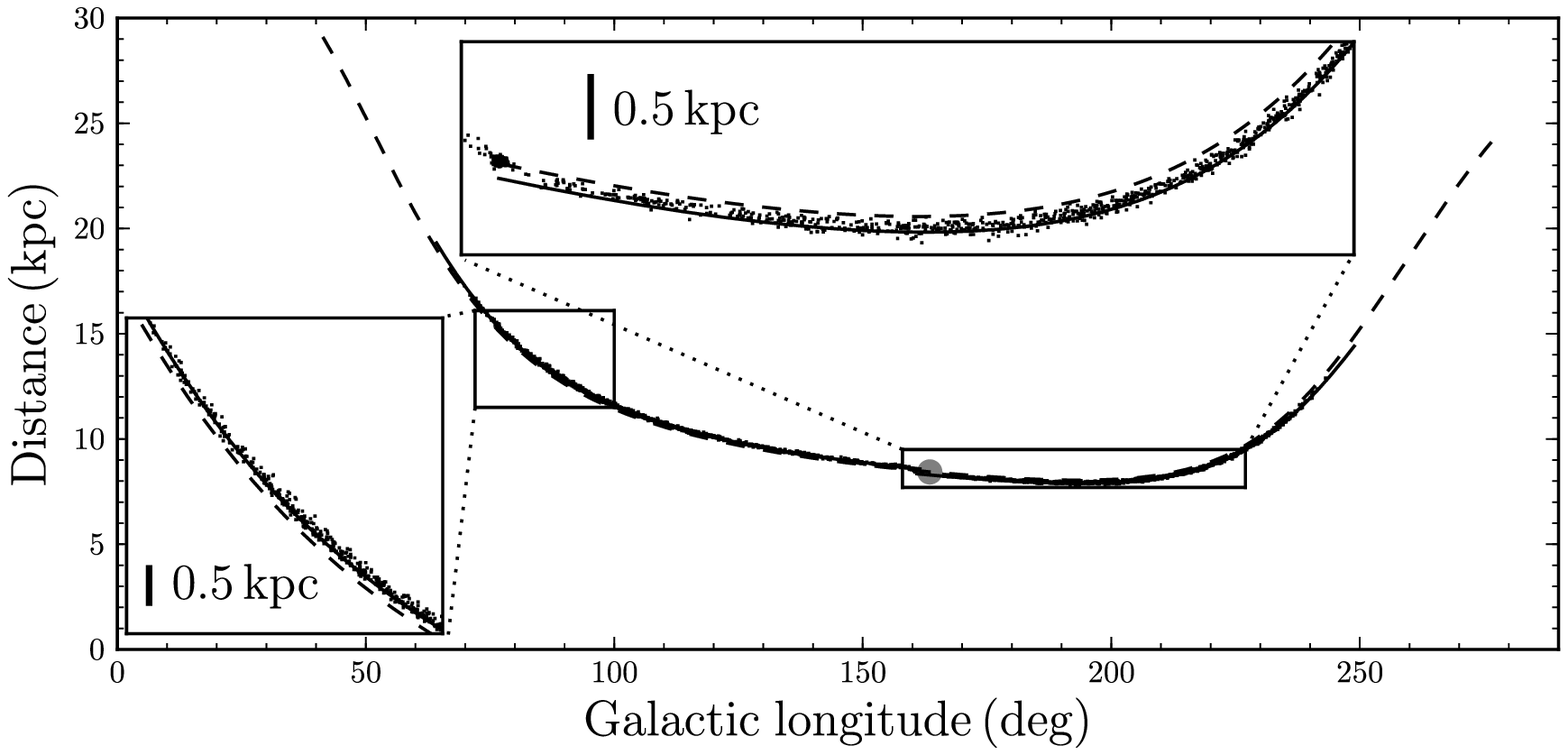}
  \caption{Simulated stream after $5.011\Gyr$ as a function of
    Galactic longitude. The top panel shows the stream in Galactic
    latitude and the bottom panel shows it in the distance from the
    Sun. The progenitor is shown as the large gray dot and its orbit
    as the dashed lines. The predicted track of the stream is shown as
    the solid line. Insets show various zoomed-in parts of the stream
    to show the stream, progenitor orbit, and predicted stream track
    up close.}\label{fig:gd1_lbd}
\end{center}
\end{figure*}

We can then estimate the dispersion around the stream track, which is
useful for getting a sense of the width of the model stream and to
find appropriate integration intervals when marginalizing the stream
PDF as discussed in \sectionname~\ref{sec:pdf}. At a given angle
offset $\Delta \apar$ we calculate the dispersion in $\Delta \opar$
from \equationname~(\ref{eq:poparapar}) and we substitute this
appropriately in $V_\veco$. We can also calculate the dispersion in
$\Delta \aperp$ using \equationname~(\ref{eq:paperpapar}). How to
estimate the dispersion in $\Delta \apar$ near a given $\Delta \apar$
is more difficult, and I simply use a dispersion of $1$, as the stream
typically spreads out over $\approx 1$ radian. We could calculate the
correlation between $\Delta \operpii$ and $\Delta \aperpii$ using
similar equations as those given in the previous paragraphs, but for a
simple estimate we can approximate these as $0.5$; I set the
correlation between $\Delta \opar$ and $\Delta \apar$ to zero. While
these estimates are not perfect, they give a reasonable estimate of
the dispersion along the stream track. This provides an adequate
starting point for more precise calculations of the dispersion in
\sectionname~\ref{sec:pdf}.

\subsection{In position--velocity coordinates}\label{sec:trackxv}

Having estimated the track of the stream and the dispersion around
this in frequency--angle space, we can propagate the track to
position--velocity coordinates by inverting the transformation
$(\vecx,\vecv) \rightarrow (\veco,\veca)$. I do this by linearizing
the transformation in the vicinity of the stream along an estimate of
the stream track, obtained using the procedure below, and inverting
this linearized transformation. That this works relies on the stream
being relatively cold (in that $\sigv/V_c \ll 1$) for a few different
reasons. First, the track of a cold stream, while in general offset
from the orbit of the progenitor, does not stray too far from it close
to the progenitor, such that the transformation to frequency--angle
coordinates can be linearized over the range of the stream--orbit
offset close to the progenitor. Second, cold streams are essentially
one-dimensional, that is, they only spread significantly over a single
direction. This means that non-linearity in the $(\vecx,\vecv)
\rightarrow (\veco,\veca)$ transformation only affects a single
direction. Therefore, we can linearize the $(\vecx,\vecv) \rightarrow
(\veco,\veca)$ transformation along a one-dimensional grid of points,
rather than on a full six-dimensional grid. Third, the coldness of the
stream means that any significant mass of the full stream PDF is close
enough to the stream track that all frequency--angle calculations can
be performed using the linear approximation along the progenitor
orbit.

\begin{figure*}[tp!!]
\begin{center}
  \includegraphics[width=0.8\textwidth,clip=]{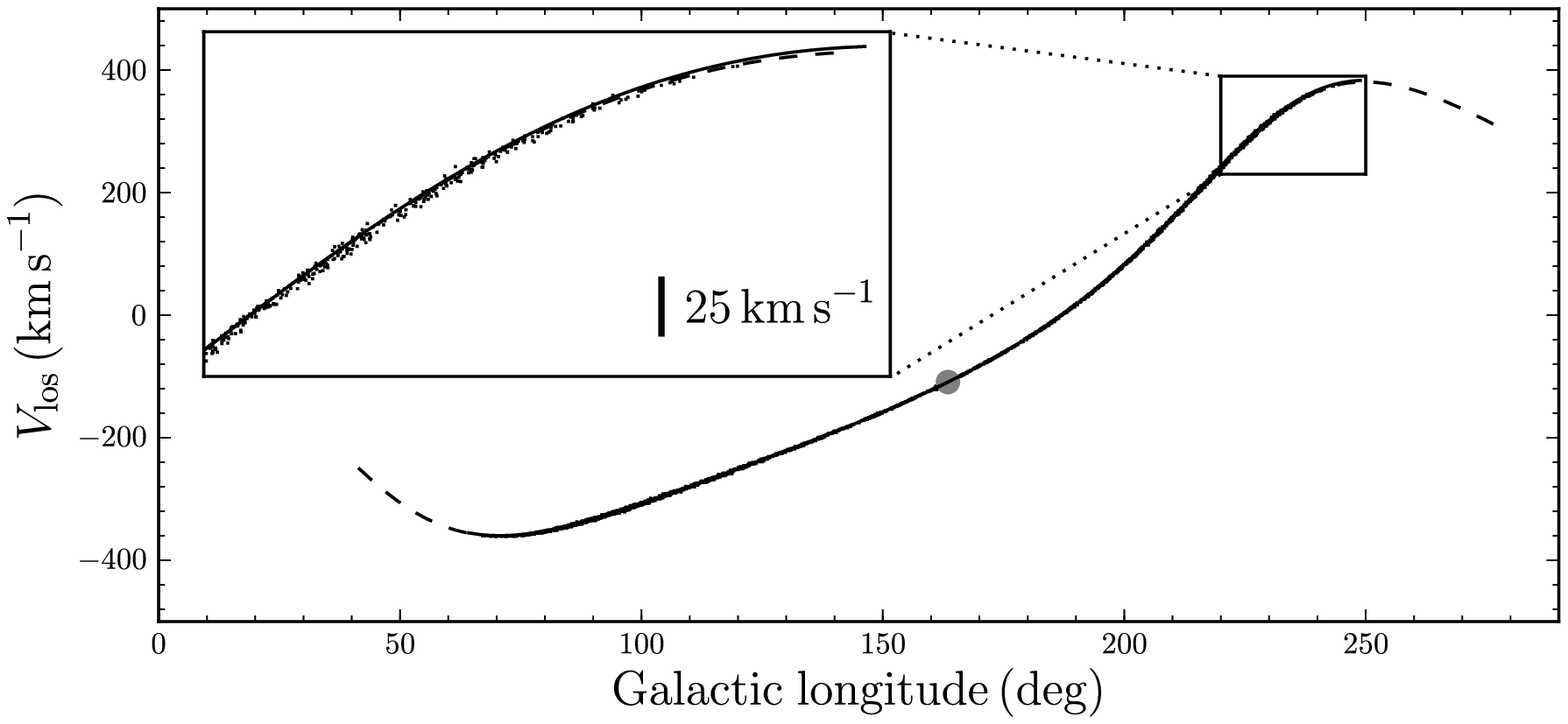}
  \includegraphics[width=0.8\textwidth,clip=]{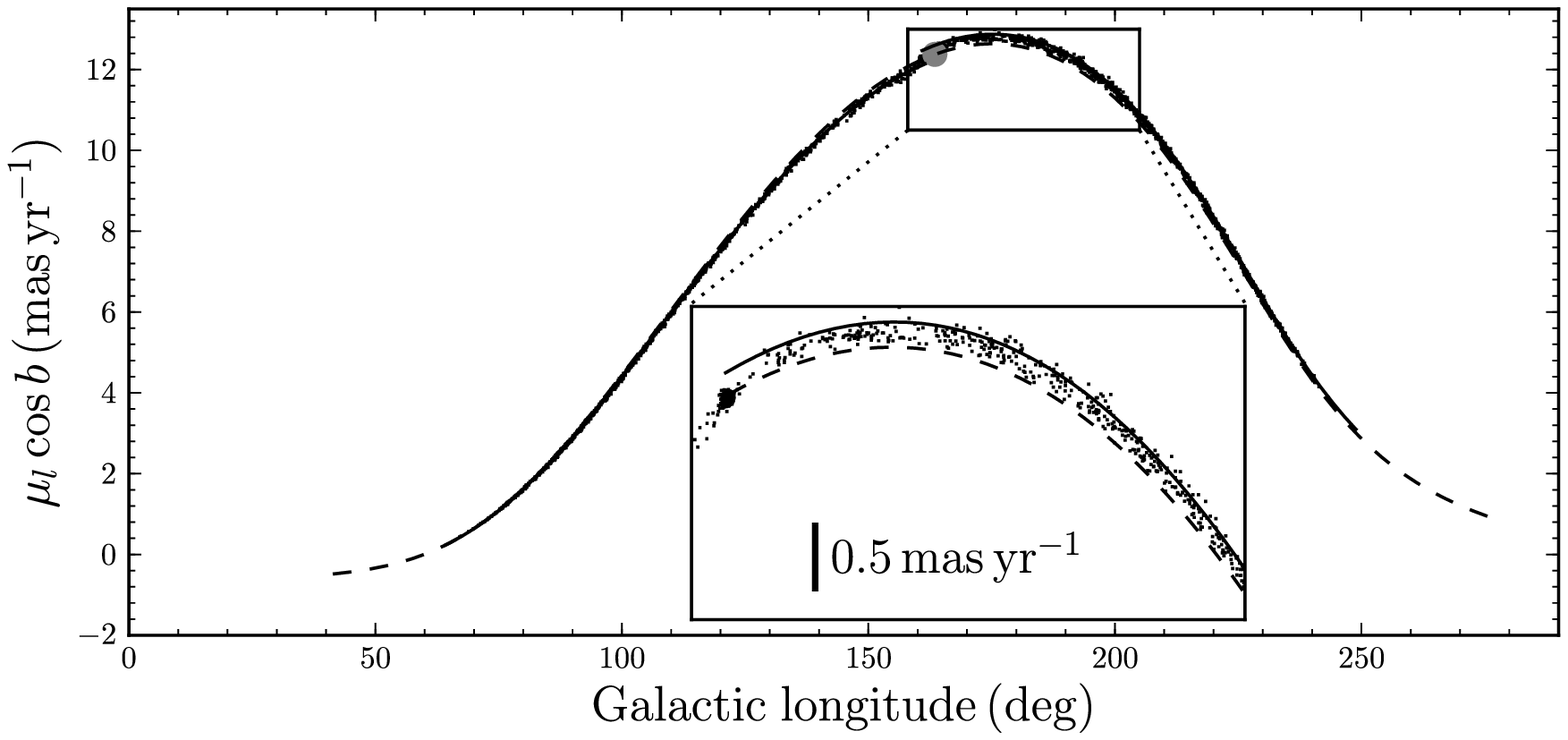}
  \includegraphics[width=0.8\textwidth,clip=]{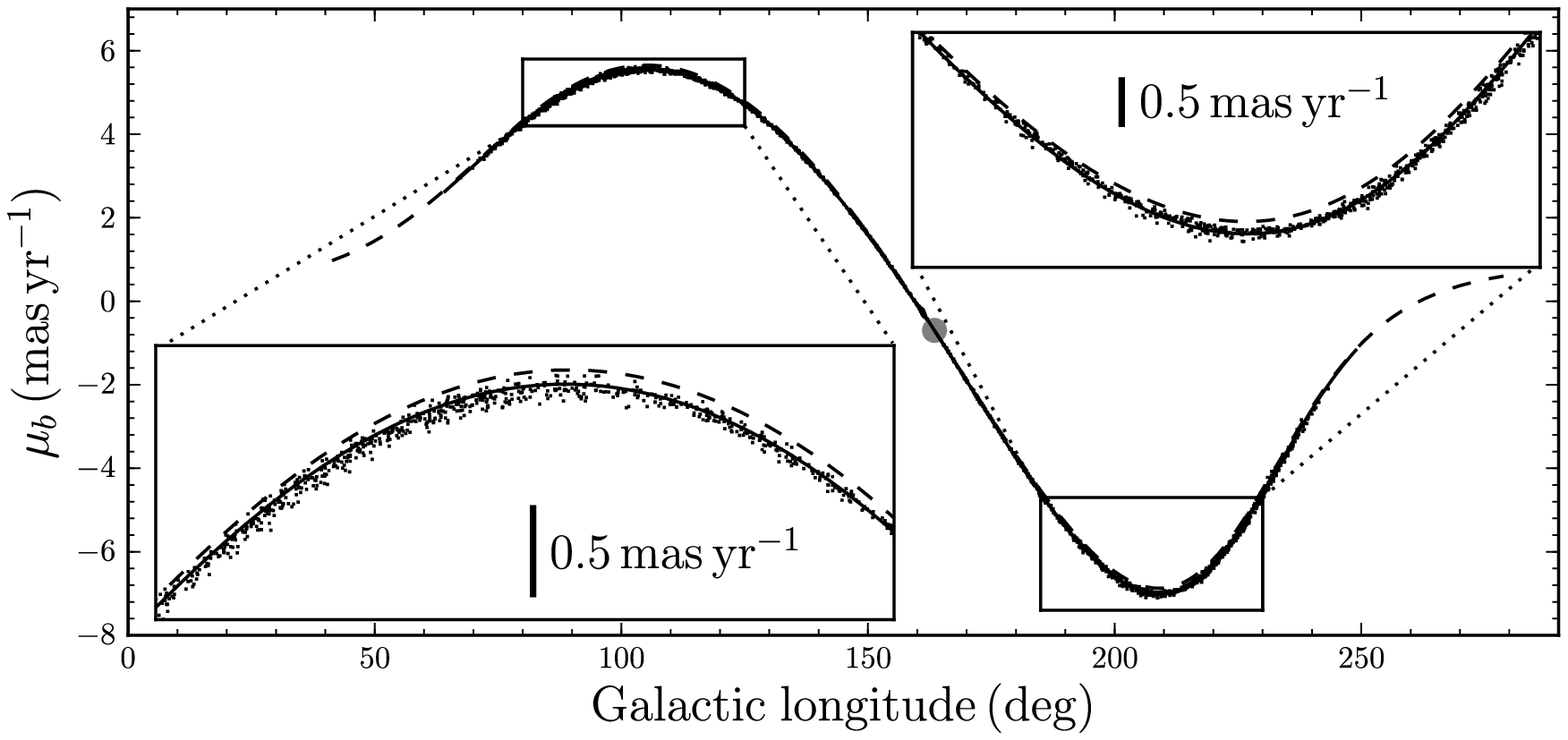}
  \caption{Simulated stream velocities as a function of Galactic
    longitude. Same as \figurename~\ref{fig:gd1_lbd}, but for the
    stream in heliocentric line-of-sight velocity $V_{\mathrm{los}}$
    (top panel), proper motion in the direction of Galactic longitude
    (middle panel), and in the direction of Galactic latitude (bottom
    panel). Insets again show various interesting parts of the stream
    up close.}\label{fig:gd1_lbv}
\end{center}
\end{figure*}

\figurename~\ref{fig:track} illustrates how I linearize the
$(\vecx,\vecv) \rightarrow (\veco,\veca)$ transformation and use it to
propagate the track of the stream to position--velocity coordinates
and to observable quantities $(l,b,D,\vlos,\pmll,\pmbb)$. The top
panel shows the orbit of the progenitor, an auxiliary orbit, and the
mean stream track in frequency--angle coordinates. Only the projection
onto the parallel direction is shown here. The auxiliary orbit is
determined by converting the mean stream frequency
$\veco=\veco_p+\Delta \mathbf{\Omega}$, where $\Delta \mathbf{\Omega}$
is determined using the techniques from the previous section, at zero
angle difference with respect to the progenitor to $(\vecx,\vecv)$ and
integrating this orbit. The transformation to configuration space is
done by linearizing the $(\vecx,\vecv) \rightarrow (\veco,\veca)$
transformation around $(\vecx_p,\vecv_p)$ and inverting this
transformation. At a small number of points along the auxiliary orbit,
chosen here to span $1.5$ radians, we calculate the Jacobian $\partial
(\veco,\veca) / \partial (\vecx,\vecv)$. In this Figure this is done
at four points, but for all other figures and calculations in this
paper eleven points are used. Each Jacobian calculation requires seven
frequency--angle calculations, such that the total number of such
computations is less than $100$ to adequately model the stream (each
computation involves a single orbit integration, see
\appendixname~\ref{sec:aa}).

We then calculate the stream track in Galactocentric
position--velocity coordinates at the parallel angles for which we
have linearized the frequency--angle transformation. This is shown in
the middle panel of \figurename~\ref{fig:track}. Similarly, we
transform the approximated variance described at the end of the
previous section to position--velocity coordinates under the linear
approximation. We can then interpolate the track of the stream between
this small number of points at which it is calculated; this gives the
full track of the stream. In \sectionname~\ref{sec:pdf}, we require an
estimate of the dispersion around the track at any angle along the
stream (for calculating the marginalized PDF). We can obtain this from
interpolating the estimated variances at the calculated track
points. This interpolation can be performed practically as follows. We
decompose the variance matrix at each calculated track point into its
eigendecomposition and order the eigenvalues by size. We then
interpolate each of the six eigenvalues using spline
interpolation. The direction of each eigenvector is interpolated using
\emph{slerp} \citep{Shoemake85a}, which is a type of spherical linear
interpolation. The variance matrix at interpolated track points can
then be constructed from the interpolated eigenvalues and
eigenvectors. All other $(\vecx,\vecv) \leftrightarrow (\veco,\veca)$
calculations can then be performed by using the closest interpolated
track point in $(\vecx,\vecv)$ or $(\veco,\veca)$ and using the
calculated Jacobian from the closest calculated track point.

Finally, we can calculate the stream track in observable quantities
$(l,b,D,\vlos,\pmll,\pmbb)$ by converting the interpolated track in
$(\vecx,\vecv)$ to these coordinates. The dispersion at calculated
track points is calculated from that in $(\veco,\veca)$ using the
appropriate Jacobians and the six-dimensional dispersion can again be
interpolated using the eigendecomposition. This is illustrated in the
bottom panel of \figurename~\ref{fig:track}.

This procedure for propagating the stream track from frequency--angle
space to configuration space works well when the misalignment between
the stream track and a single orbit is not too large (as is the case
for the example used throughout in this paper). However, when this
misalignment is large, the auxiliary orbit deviates significantly from
the stream track and the linear approximations to go from auxiliary
orbit to stream track break down. This can be diagnosed by calculating
the frequencies and angles along the estimated stream track in
configuration space using the algorithm in \appendixname~\ref{sec:aa}
and comparing it to those of the desired stream track computed using
the methods from the previous section. When the misalignment is large
and the linear approximations break down, these two do not agree (in
that they deviate by much more than the spread in the stream around
the mean track).

We can fix this by iterating the calculation of the stream
track. After the first estimate of the stream track is obtained using
the auxiliary orbit, we can compute another estimate starting from the
previous estimate of the stream track---now the auxiliary
\emph{track}, because it is no longer a single orbit---in
$(\vecx,\vecv)$ in the same way as the first estimate was
calculated. Thus, we linearize the $(\vecx,\vecv) \rightarrow
(\veco,\veca)$ transformation around points along the auxiliary track
and calculate a new estimate of the stream track based on the offset
in frequency--angle between the auxiliary track and the desired
track. Even for large misalignments, this procedure converges in a few
iterations (that is, the difference between the calculated stream
track and the desired stream track in frequency--angle becomes much
smaller than the dispersion within the stream).

The average stream location computed in the manner described in this
section is an approximate track. The mean stream location, for
example, in distance from the Sun at a given Galactic longitude, can
be exactly calculated by marginalizing over the full six-dimensional
stream PDF in $(l,b,D,\vlos,\pmll,\pmbb)$ over the unobserved
dimensions $(b,\vlos,\pmll,\pmbb)$. In general this will give a
slightly different stream location, because the stream DF is not
exactly Gaussian. However, it is demonstrated in
\sectionname~\ref{sec:pdf}, where I calculate the full stream PDF,
that the approximate mean stream location of this section is
sufficiently close to the true average position for all practical
purposes. The same should hold for any cold stream.

\figurename~\ref{fig:gd1_xz} shows the simulated stream particles as
well as the stream track calculated in this section in Galactocentric
$X$ and $Z$ coordinates. It is clear that the average stream location
tracks the position of the simulated stream. \figurename
s~\ref{fig:gd1_lbd} and \ref{fig:gd1_lbv} show the same, but in
observed coordinates $(l,b,D,\vlos,\pmll,\pmbb)$. These figures show
that the simple fiducial stream model does a good job of predicting
where the stream is located as a function of Galactic longitude.

\section{Mock stream data}\label{sec:mock}

\begin{figure}[t!]
 \includegraphics[width=0.48\textwidth,clip=]{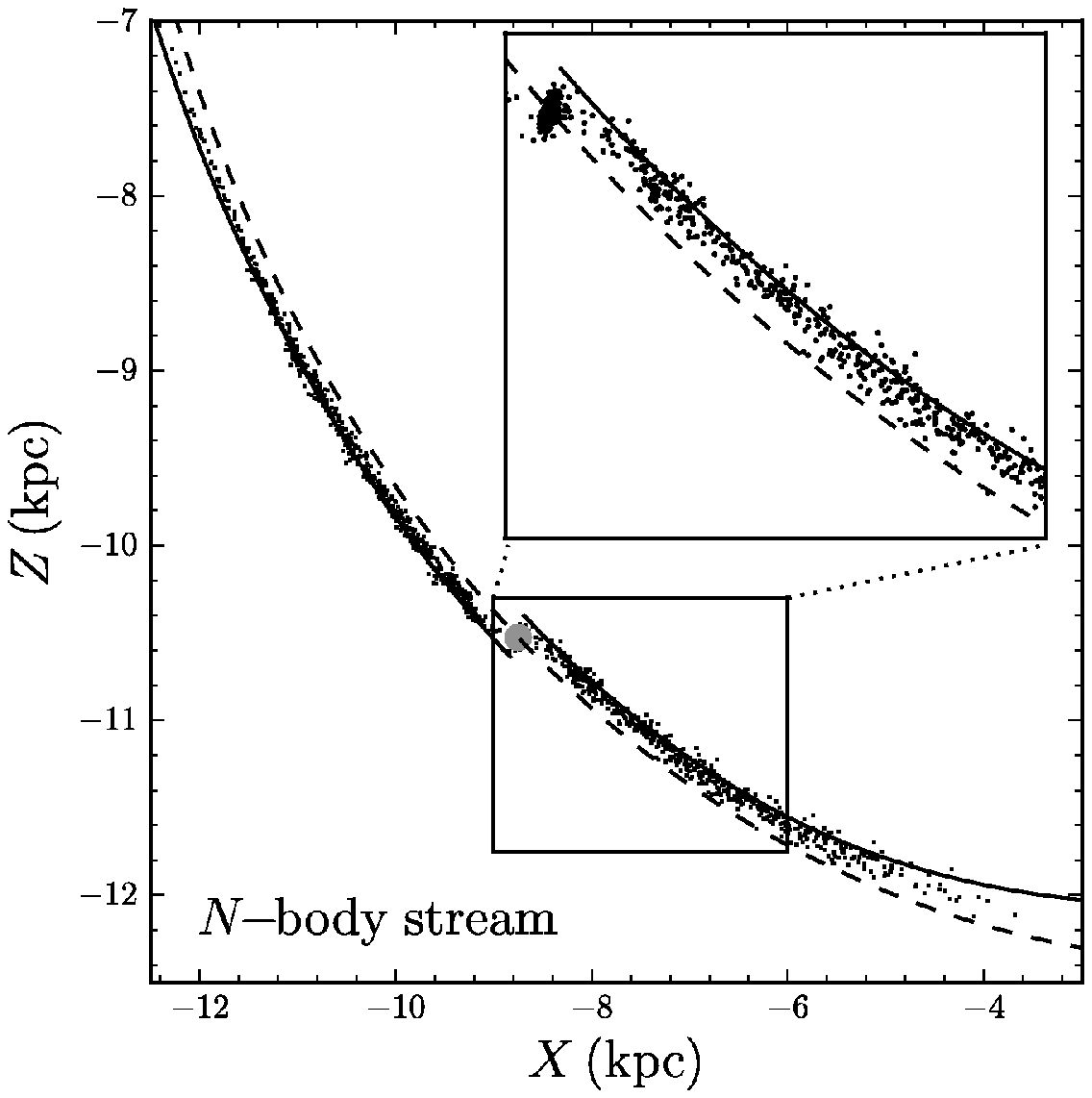}\\
 \includegraphics[width=0.48\textwidth,clip=]{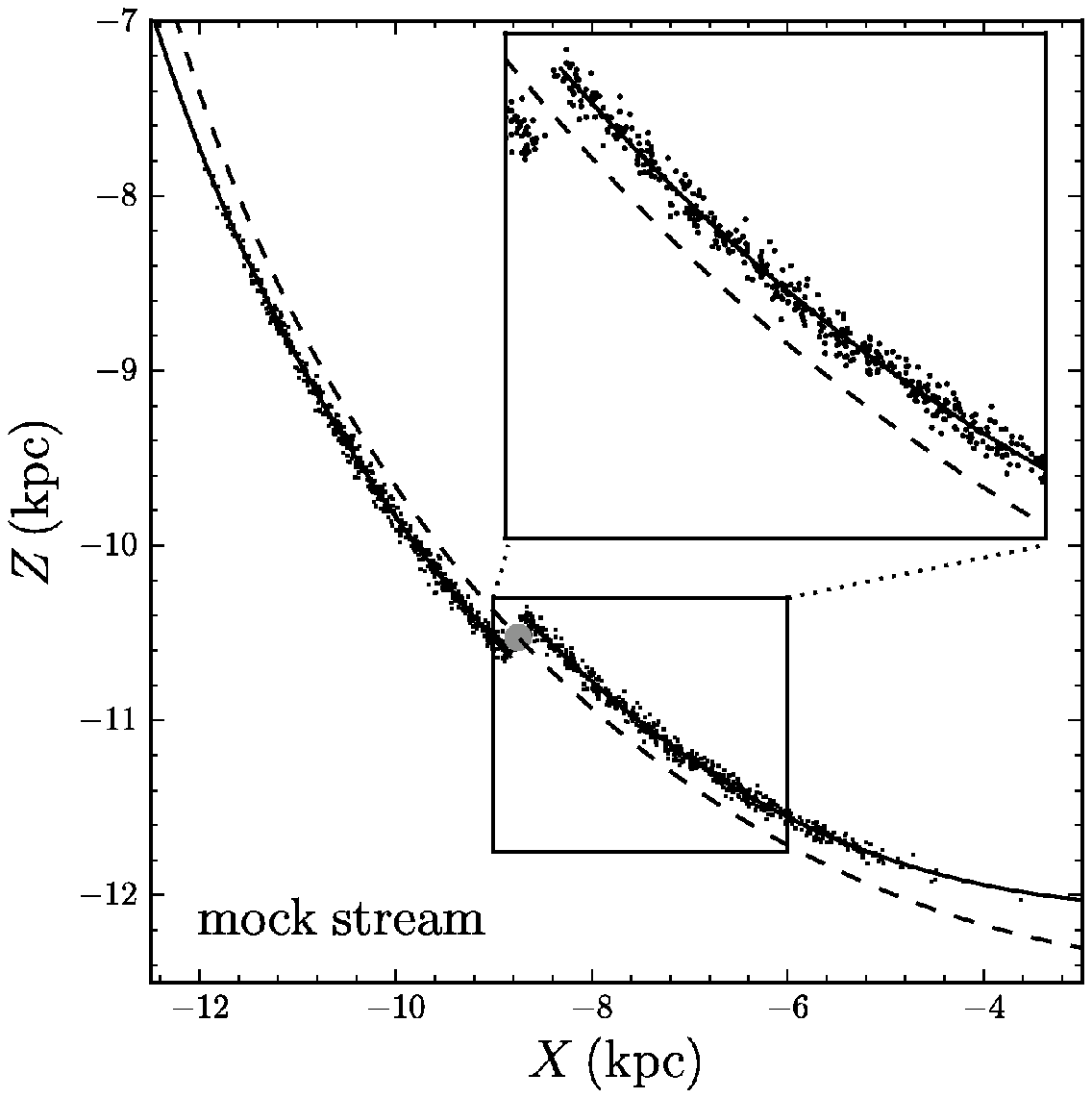}
  \caption{Mock stream data generated from the model. The top panel
    shows the same $N$-body stream as shown in previous figures, but
    $1.96\Gyr$ earlier in its evolution. Symbols, lines, and the
    stream model are the same as those in
    \figurename~\ref{fig:gd1_xz}, except that the model progenitor
    phase-space location from \figurename~\ref{fig:gd1_xz} has been
    integrated backward in time for $1.96\Gyr$ and the model
    disruption time was reduced by $1.96\Gyr$ (no other changes were
    made to the model's parameters). The bottom panel shows mock stream
    data drawn from the model using the procedures discussed in
    \sectionname~\ref{sec:mock}. The gap between the leading and the
    trailing arm is due to the difference in mean frequency between
    the arms. Overall, the spatial distribution of the mock data is
    very similar to that of the $N$-body stream.}\label{fig:gd1_mock}
\end{figure}

Using the generative model described in
\sectionname~\ref{sec:modeloa}, it is straightforward to draw mock
stream data in frequency--angle space. The three ingredients---time,
frequency, and angle distributions of the debris---allow for mock data
to be generated in four steps. First, a stripping time $t_s$ is
sampled from the distribution of stripping times. Second, a frequency
offset $\Delta \veco$ with respect to the progenitor is drawn from the
frequency distribution at $t_s$. Third, an initial angle offset is
drawn. Fourth, the initial angle offset is incremented by $\Delta
\veco \,t_s$. This generates the final frequency--angle coordinates of
the mock stream member $(\Delta \veco,\Delta \veca+\Delta \veco\,t_s)
+ (\veco_p,\veca_p)$.

For the fiducial stream model of \sectionname~\ref{sec:fidmodel} the
procedure for producing mock data is simplified to sampling a small
number of easy-to-sample distributions. First, $t_s$ is drawn
uniformly between $0$ and $t_d$. Second, $\Delta \operp$ is drawn from
the Gaussian distribution of $\Delta \operp$. The distribution of
$\Delta \opar$ is not Gaussian, as it is a Gaussian in $|\Delta
\opar|$ multiplied with $|\Delta \opar|$ over the range $\Delta \opar
> 0$. This distribution is log-concave, that is, the derivative of its
logarithm is negative everywhere, and therefore it can be efficiently
sampled using adaptive-rejection sampling \citep{Gilks92a}. The sign
of $\Delta \opar$ is determined by whether we are generating a leading
or trailing stream. Third, initial angle offsets are drawn from the
Gaussian distribution of such offsets. The fourth step is as above.

To transform the mock stream in frequency--angle coordinates to
position--velocity space, we use the approximate procedure for this
transformation along the stream track described at the end of the
previous section. The mock stream in $(\vecx,\vecv)$ can then be
further transformed to observable coordinates
$(l,b,D,\vlos,\pmll,\pmbb)$ or any other similar quantities as
desired.

\figurename~\ref{fig:gd1_mock} shows mock data generated for the model
of the simulated stream described in
\sectionname~\ref{sec:fidmodel}. The mock data in this Figure have
been generated $1.96\Gyr$ before the final snapshot of the simulation
(that is, $1.96\Gyr$ before \figurename~\ref{fig:gd1_xz}). The model
from \sectionname~\ref{sec:fidmodel}, which was for the final
snapshot, was adjusted to this earlier time by backward orbit
integration of the model progenitor and by revising the disruption
time downward by $1.96\Gyr$. No other changes were made to the
model. The bottom panel of \figurename~\ref{fig:gd1_mock} shows the
mock data generated from the model in $(X,Z)$, while the top panel
shows the simulated stream data at this time. Overall, the
distribution of the mock and simulated data are similar.

\section{The full stream PDF}\label{sec:pdf}

Data on tidal streams typically come in one of two flavors: data on
individual stream members, often with multiple missing phase--space
components, or data describing the mean position and width of the
stream. For the proper analysis of these kinds of data it is useful to
be able to evaluate the PDF for the stream model described in this
paper and to marginalize over or condition it on certain dimensions.

The generative model described in \sectionname~\ref{sec:modeloa}
corresponds to the stream PDF in frequency $\veco$, angle $\veca$, and
stripping time $t_s$
\begin{equation}
  p(\veco,\veca,t_s) = p(t_s)\,p(\veco|t_s)\,p(\veca|\veco,t_s)\,,
\end{equation}
where the three factors on the right are specified through the three
ingredients of \sectionname~\ref{sec:modeloa}. The PDF in
position--velocity space is then given by
\begin{equation}\label{eq:pxvt}
  p(\vecx,\vecv,t_s) = p(\veco,\veca,t_s)\,\left|\frac{\partial
    \veco}{\partial \vecj}\right|\,,
\end{equation}
where $|\partial \veco / \partial \vecj|$ is equal to the Jacobian
$|\partial (\veco,\veca)/\partial(\vecx,\vecv)|$ because $|\partial
(\vecj,\veca)/\partial(\vecx,\vecv)|=1$.  The PDF in observable
quantities $(l,b,D,\vlos,\pmll,\pmbb)$ can be obtained from
$p(\vecx,\vecv,t_s)$ by multiplying by the Jacobian
$\left|\partial(\vecx,\vecv)/\partial(l,b,D,\vlos,\pmll,\pmbb)\right|$.

\begin{figure*}[t!]
 \includegraphics[width=0.32\textwidth,clip=]{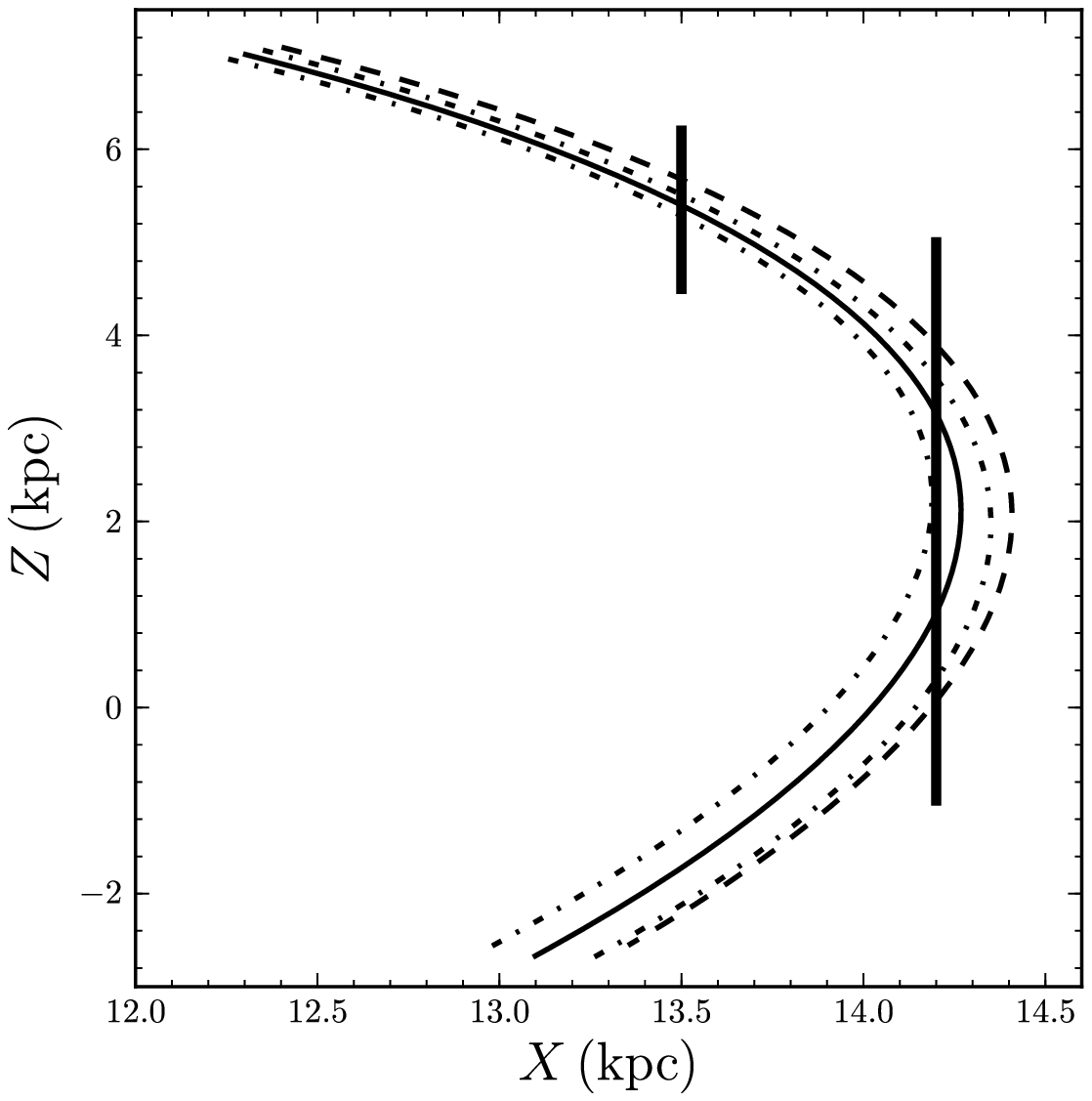}
 \includegraphics[width=0.32\textwidth,clip=]{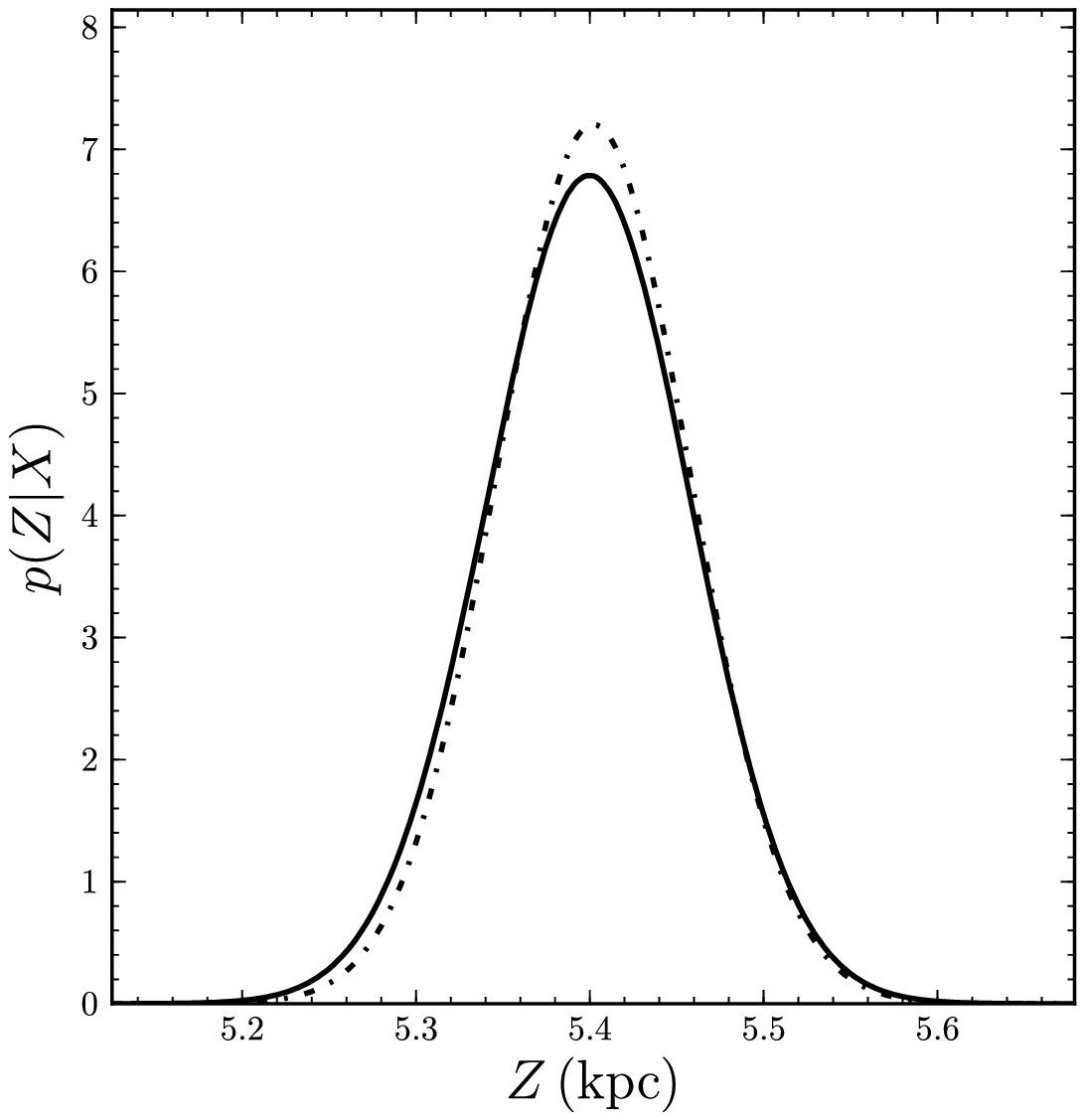}
 \includegraphics[width=0.32\textwidth,clip=]{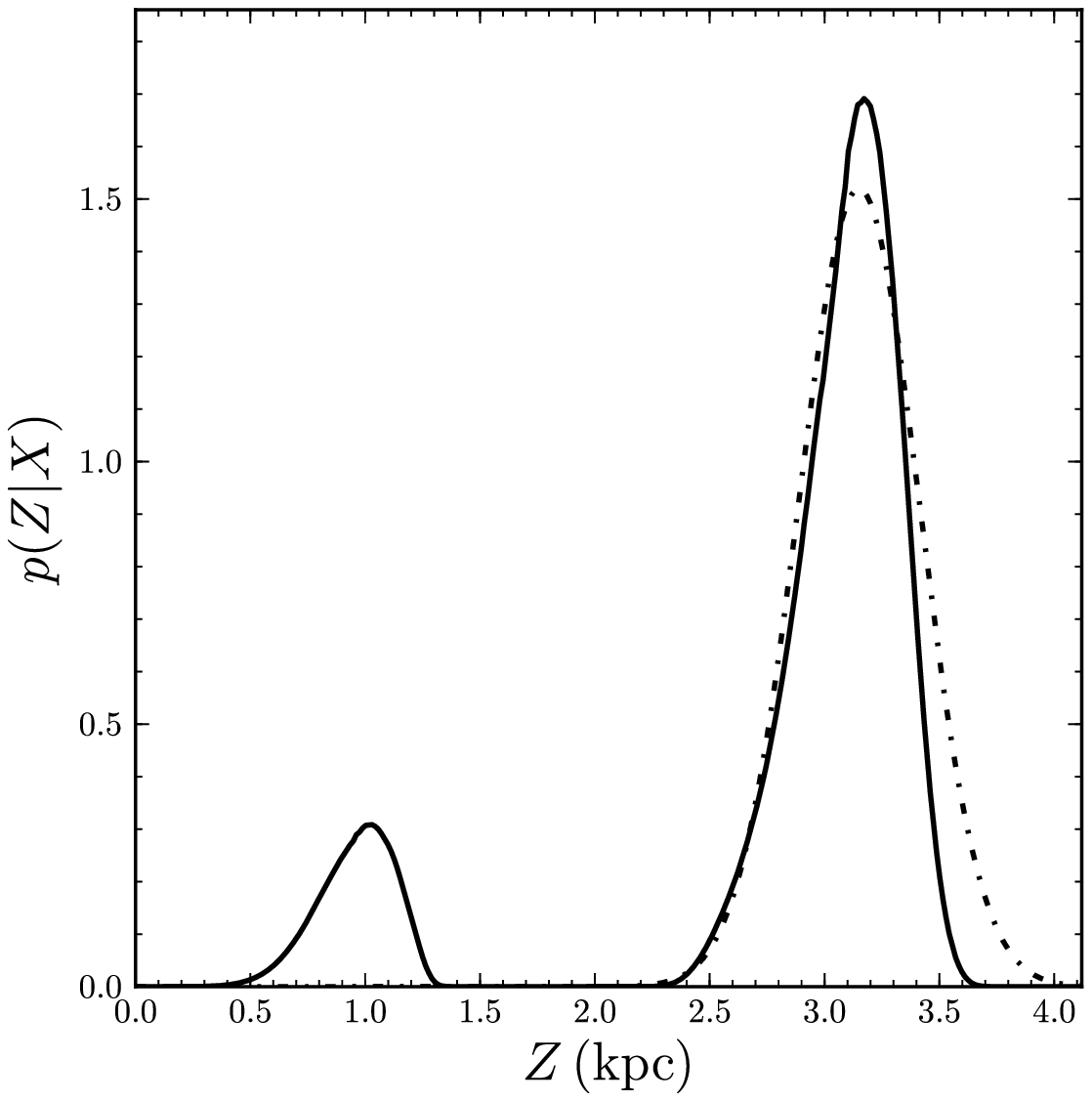}
 \caption{Marginalized, conditional $p(Z|X)$ PDFs. The left panel
   shows the model leading stream (see \figurename~\ref{fig:gd1_xz}),
   with the stream track shown as the solid line, the estimated
   $2\sigma$ width of the stream as the dash-dotted line, and the
   progenitor orbit as the dashed line. The solid line in the middle
   and right panels shows $p(Z|X)$ calculated at the $X$ indicated in
   the left panel by thick vertical lines; the dash-dotted line is a
   simple Gaussian estimate. The upper branch near $Z \approx 5\kpc$
   of the stream is chosen for the middle panel. The right panel shows
   an $X$ close to where the stream turns around in $X$; the PDF has
   multiple peaks in this case.}\label{fig:gd1_pdf_xz}
\end{figure*}

The stripping time $t_s$ cannot be observed and should therefore be
marginalized over when evaluating the PDF. One of the great advantages
of modeling a stream in frequency--angle coordinates is that this
marginalization can be calculated analytically for many choices of
$p(t_s)$ if $p(\veco|t_s) \equiv p(\veco)$. Specifically, for the
fiducial model $p(t_s)$ is uniform up to a maximum disruption time
$t_d$ (see \equationname~[\ref{eq:pt}]) such that we can write
\begin{equation}
\begin{split}
  p(& \veco,\veca)\\
  & = \int \dd t_s\,p(\veco,\veca,t_s)\,,\\
  & = p(\veco)\,\int_0^{t_d} \dd t_s\,
  \frac{1}{(2\,\pi)^{3/2}\,\sigma_\theta^3}\\
  & \qquad \qquad \times\exp\left(-\frac{1}{2\,\sigma_\theta^2}\left[\sum_{R,\phi,Z}\left(\Delta \veca_i-\Delta \veco_i\,t_s\right)^2\right]\right)\,,
\end{split}
\end{equation}
where the sum in the argument of the exponential is over the three
components of $\Delta \veca_i$ and $\Delta \veco_i$. This integral can
be done analytically, resulting in
\begin{equation}\label{eq:pdf}
\begin{split}
  p(& \veco,\veca)  = 
  p(\veco) \,  \frac{\mathrm{erf}(a_0) + \mathrm{erf}(a_d)}{4\,\pi\,\sigma_\theta^2\,|\Delta \veco|}\\
  & \ \times \exp\left(-\frac{1}{2\,\sigma_\theta^2}\left[\sum_{R,\phi,Z}\Delta \veca^2_i-\frac{\left(\sum_{R,\phi,Z} \Delta \veco_i\Delta \veca_i\right)^2}{|\Delta \veco|^2}\right]\right)\,.
\end{split}
\end{equation}
In this expression, $\Delta X^2 \equiv (X_1-X_2)^2$. The expression in
square brackets is positive by using the Cauchy-Schwarz
inequality. The quantities $a_0$ and $a_d$ in this expression are
defined by the following expressions
\begin{align}
  \tilde{t}_s & = \frac{\sum \Delta \veco_i\Delta \veca_i}{|\Delta \veco|^2}\,,\\
  a_0 & = \frac{|\Delta \veco|}{\sqrt{2}\,\sigma_\theta}\,\tilde{t}_s\,,\\
  a_d & = \frac{|\Delta \veco|}{\sqrt{2}\,\sigma_\theta}\,\left(t_d - \tilde{t}_s\right)\,.
\end{align}
The time $\tilde{t}_s$ is the best estimate of the stripping time for
a given $\Delta \veco$ and $\Delta \veca$. The quantity $a_0$ then
serves to suppress the PDF in case the stripping time is smaller than
zero, \ie, a stream member appears to have been removed \emph{in the
  future}; in this case $\mathrm{erf}(a_0) \approx -1$ and the PDF
goes to zero. Similarly, $a_d$ suppresses the PDF when the stripping
time is larger than $t_d$, \ie, the stream member seems to have been
stripped before disruption began. Both of these get a tolerance
corresponding to the initial angle spread. For example, if
$\tilde{t}_s$ is negative, but so small such that $|\Delta
\veco|\,\tilde{t}_s \sim \sigma_\theta$, the PDF is only mildly
suppressed. For stars well within the stream $\mathrm{erf}(a_0) +
\mathrm{erf}(a_d) \approx 2$.

We can marginalize \equationname~(\ref{eq:pxvt}) over $t_s$ and find
that
\begin{equation}\label{eq:pxv}
  p(\vecx,\vecv) = p(\veco,\veca)\,\left|\frac{\partial
    \veco}{\partial \vecj}\right|\,.
\end{equation}
When evaluating this PDF as a function of $(\vecx,\vecv)$ or as a
function of observable quantities $(l,b,D,\vlos,\pmll,\pmbb)$ we
calculate frequencies and angles using the approximate, linear
transformation near the stream track as described at the end of
\sectionname~\ref{sec:trackxv}. Thus, $p(\vecx,\vecv)$ can be
evaluated very quickly by making efficient use of array operations. In
this approximation the Jacobian $|\partial \veco / \partial \vecj|$ is
constant everywhere near the track (for a given potential).

\begin{figure*}[t!]
\begin{center}
 \includegraphics[width=0.8\textwidth,clip=]{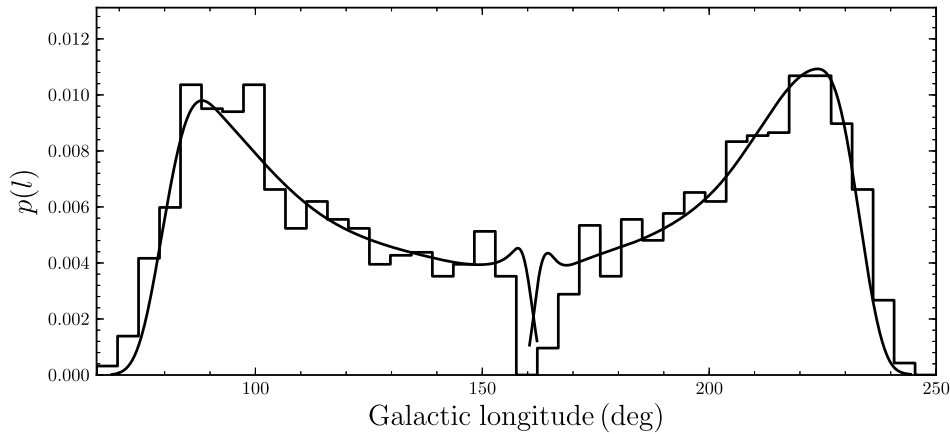}
 \caption{Density distribution as a function of Galactic
   longitude. The smooth solid line shows the fiducial model's
   prediction, while the histogram shows the density along the
   simulated stream. The simple fiducial stream model can accurately
   predict the density distribution along the
   stream.}\label{fig:gd1_pdf_l}
\end{center}
\end{figure*}

\begin{figure}[t!]
  \includegraphics[width=0.48\textwidth,clip=]{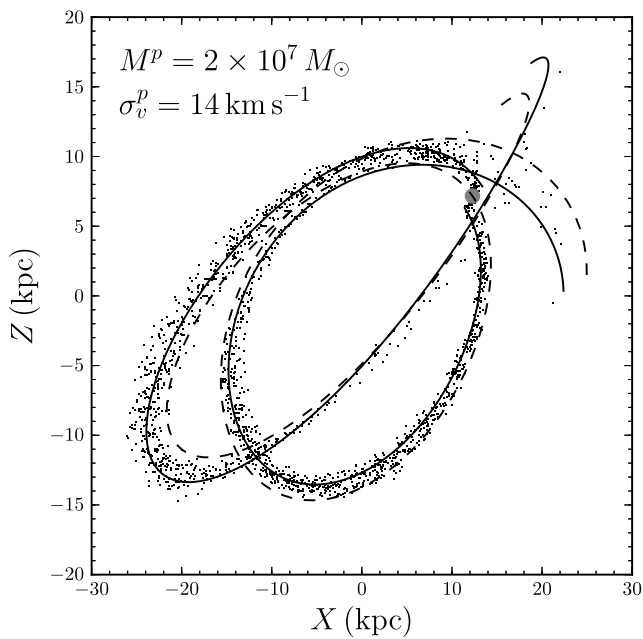}
  \caption{Same as \figurename~\ref{fig:gd1_xz}, but for a simulation
    of a progenitor with a mass $M^p$ that is $1,000$ times larger and
    therefore a velocity dispersion $\sigma_v^p$ that is ten times
    larger. The model used to predict the stream track (solid line) is
    the same as that used in \figurename~\ref{fig:gd1_xz}, except that
    the model velocity dispersion \sigv\ has been increased by a
    factor of ten. The stream spreads over a much larger volume for a
    heavier progenitor, but the fiducial model still accurately
    predicts its location in phase space.}\label{fig:gd1-hisigv_xz}
\end{figure}

We can marginalize the PDF over unobserved directions, convolve it
with uncertainty distributions, or condition it on certain directions
using the standard rules of probability theory; I will refer to all of
these as marginalizations in what follows, as they all involve
integrations over the PDF. In practice, to perform these
marginalizations it is useful to estimate the extent of the PDF to
define appropriate intervals for efficient numerical integration. For
this we can use the average stream track and dispersion around it as
estimated in \sectionname~\ref{sec:track}. This is most easily
explained through an example. To determine $p(Z|X)$, the distribution
of Galactocentric $Z$ at a given Galactocentric $X$, we need to
calculate
\begin{equation}
  p(Z|X) = \frac{p(Z,X)}{p(X)} = \frac{\int \dd Y\,\dd \vecv\, p(\vecx,\vecv)}{\int \dd Y\,\dd Z\,\dd \vecv\, p(\vecx,\vecv)}\,.
\end{equation}
Thus, we need to marginalize over $(Y,\vecv)$ in the numerator and
$(Y,Z,\vecv)$ in the denominator. Focusing on the numerator, we find
the closest point on the stream track calculated as in
\sectionname~\ref{sec:trackxv} and calculate its six-dimensional
Gaussian approximation. Then we condition this six-dimensional
Gaussian on $(X,Z)$ using the standard rules of Gaussian conditioning
(\eg, Appendix B of arXiv version 1 of \citealt{Bovy11a}) to obtain
the approximate PDF $p(Y,\vecv|X,Z)$. We then evaluate $\int \dd
Y\,\dd \vecv\, p(\vecx,\vecv)$ by numerical integration over the, \eg,
$3\sigma$ range of this Gaussian; this numerical integration can be
performed efficiently using Gaussian quadrature. 

\figurename~\ref{fig:gd1_pdf_xz} shows such PDFs of $p(Z|X)$ for a few
$X$ on the leading arm of the stream as well as the estimated Gaussian
PDFs determined using the procedure in
\sectionname~\ref{sec:trackxv}. It is clear that the average stream
location as calculated in \sectionname~\ref{sec:trackxv} is very close
to the actual average location of the stream in $(\vecx,\vecv)$ and
that the estimated dispersion is close to the true model
dispersion. The fact that the estimated Gaussian in the right panel is
only at the highest peak is by choice; there is an estimated Gaussian
dispersion at the lower peak as well, but it is not shown.

Similar procedures can be followed to evaluate other projections and
marginalizations of the PDF (\eg, $p(b|l)$ for the observed track of
the stream on the sky), to convolve the PDF with observational
uncertainties, and to calculate moments of the PDF, \eg, the density
and velocity dispersion along the stream. Another example is shown in
\figurename~\ref{fig:gd1_pdf_l}. This Figure shows the model's density
as a function of Galactic longitude and compares it to that of the
simulated stream in the final snapshot (\ie, that of \figurename
s~\ref{fig:gd1_lbd} and \ref{fig:gd1_lbv}). The model density here is
calculated using a Monte Carlo mock stream sample, drawn as described
in \sectionname~\ref{sec:mock}, rather than through direct numerical
integration.

\section{Discussion}\label{sec:discussion}

\subsection{At what progenitor mass does the framework break down?}

I have illustrated the new framework for modeling tidal streams using
a cold, narrow stream with a progenitor velocity dispersion of only
$\sigma_v^p = 1.4\kms$ (and a progenitor mass of $M^p = 2 \times
10^4\msun$). For a tidal stream that is as cold as that, the offset
between the progenitor's orbit and the stream track as well as the
dispersion within the stream are small (we expect these relative
offsets to be approximately $\sigv/V_c$). Therefore, we can linearize
the $(\vecx,\vecv) \leftrightarrow (\veco,\veca)$ transformation over
the relevant range of $(\vecx,\vecv)$.  The stream dispersion and the
stream--progenitor-orbit offset scale approximately as mass$^{1/3}$,
such that for streams arising from the tidal disruption of small dwarf
galaxies, we expect the stream--progenitor-orbit offset to be around
$10\,\%$. In the fiducial model and for the simulated stream used in
this paper, the internal frequency dispersion in the stream is six
times smaller than the stream--progenitor-orbit offset. Therefore,
bridging the latter with a linearized frequency--angle transformation
is the more stringent constraint.

To test whether the framework presented applies to streams arising
from the tidal disruption of small dwarf galaxies, I have run the same
simulation as that described in \sectionname~\ref{sec:sim}, but for a
progenitor mass of $M^p=2\times 10^7\msun$ and a velocity dispersion
of $\sigma_v^p=14\kms$. The tidal radius of this progenitor is
$0.69\kpc$ and I use a softening parameter of $14\pc$ in this
case. The final snapshot of this simulation in Galactocentric $X$ and
$Z$ coordinates is shown in \figurename~\ref{fig:gd1-hisigv_xz}. I
predict the stream track for this simulated stream by taking the same
fiducial model as described in \sectionname~\ref{sec:fidmodel}, except
that the model velocity dispersion \sigv\ is increased by a factor of
ten. This track is shown in \figurename~\ref{fig:gd1-hisigv_xz}.

It is clear from \figurename~\ref{fig:gd1-hisigv_xz} that the fiducial
model as well as the action--angle approximations used in this paper
still work for heavier progenitors with masses $\approx
10^7\msun$. Similar experiments with progenitors with masses $\approx
10^8\msun$ show that the framework presented here can model the tidal
streams produced by such heavy progenitor's as well. When the
stream--progenitor-orbit becomes so large that the assumption of a
linear $(\vecx,\vecv) \leftrightarrow (\veco,\veca)$ transformation
becomes invalid, we can improve the calculation of the stream track by
iterating the procedure to calculate the stream track as described in
\sectionname~\ref{sec:trackxv} and illustrated in
\figurename~\ref{fig:track}. With this procedure, streams originating
from progenitors up to masses of $10^9\msun$ should be able to be
modeled with the framework presented in this paper.

Above $M^p = 10^9\msun$ the effects of dynamical friction, the dwarf
galaxy's gravity, and non-linearity in the $\vecj \rightarrow \veco$
transformation are expected to start affecting the structure and
location of the stream \citep[\eg,][]{Sanders13a}. Therefore, streams
originating from the heaviest dwarf satellites will have to be modeled
with more intricate means.

\subsection{Dynamical fitting}

There are multiple existing methods for fitting data on tidal streams
with models for the Milky Way potential. However, none of these allow
for the likelihood of a model to be evaluated using a well-defined
generative model
\citep[\eg,][]{Penarrubia12a,Sanders13b,PriceWhelan13a}. The only
exceptions to this are orbit-fitting methods, but these use a faulty
generative model (see below), and the method of
\citet{Varghese11a}. The latter is in many ways similar to the method
proposed here, in that it calculates the stream track and uses this as
the basis of the stream inference. They calculate the track directly
in position--velocity space by orbit integration of stream members
released at the Lagrange points. However, they do not relate the
observed width of the stream to the velocity dispersion of the
progenitor---and thus the location of the Lagrange points and the
position of the track---which is therefore only weakly constrained and
has to be assumed. In general, it is straightforward to write down
generative models of tidal streams in position--velocity space by
substituting our initial offsets $(\Delta \veco,\Delta \veca)$ by
similar offsets $(\Delta \vecx,\Delta \vecv)$ and integrating both the
progenitor and offset stream members forward in time. However,
evaluating such models when fitting observational data requires large
numbers of orbit integrations to marginalize over stripping time
(which cannot be performed analytically in this case) and missing
phase--space dimensions. Thus, any such method will likely be orders
of magnitude slower than the method put forth here.

The forward model of this paper is superior to other stream-fitting
methods in that it allows easy marginalization over missing or noisily
measured phase--space coordinates. The obvious exception to this is
orbit fitting, which as a purely one-dimensional model is easy to
integrate over. However, orbit fitting likely produces biased results
as streams do not delineate single orbits over their full angle
widths.  The fiducial model I propose in this paper allows for a
straightforward replacement of orbit fitting, at the expense of adding
two extra parameters, $\sigv$ and $t_d$, in addition to the
six-dimensional position of the progenitor. The orbit in orbit fitting
then gets replaced by the stream track, calculated as in
\sectionname~\ref{sec:track}. To constrain $\sigv$ and $t_d$ it is
necessary to also use the observed width and length of the
stream. While cold streams are typically too narrow to be resolved in
$D$ or any of the velocity components, the observed width of the
stream's sky projection can be measured and used as a
constraint. Future high-resolution spectroscopic data,
$\mu\mathrm{as}$-level astrometric data, or percent-level distances
for standard candles may allow the width and structure of tidal
streams to be resolved and provide further constraints on the
generative model. The switch from orbit fitting to stream-track
fitting comes at the expense of $\approx100$ orbit integrations for
each model rather than a single integration. If the progenitor is
unknown we also need to marginalize over whether we are seeing a
leading or a trailing stream.

The dynamical framework proposed here can be used in situations where
fitting a stream track is insufficient, for example, when high-quality
data in some dimensions on individual stars are available. Many of the
approaches in the literature do not require a model progenitor
position. This includes methods that attempt to minimize the spread of
orbital energies or actions in the stream (\eg,
\citealt{Binney08a,Penarrubia12a}, Sanderson \etal, 2014, in
preparation) as well as the method of \citet{Sanders13b}, which
constrains the potential by requiring angle differences in the stream
to lie along the same direction as frequency differences (see
\equationname~\ref{eq:pdf} and subsequent discussion). These methods
require good measurements of the phase--space coordinates of stream
members, or at the very least, \emph{some} measurement of all of the
six phase--space coordinates (although the latter can be relaxed when
excellent measurements of the line-of-sight velocity along the stream
are available; \citealt{Binney08a}). If such high-quality data are
present, these methods can provide good initial guesses for the PDF of
potential parameters that fit a stream, which can subsequently be used
in a more thorough exploration of the PDF and structure of the stream
using the framework of \sectionname~\ref{sec:pdf}. That these methods
work also goes to show that the progenitor parameters---phase--space
position, $\sigv$ and $t_d$---are not that important for the
constraints on the gravitational potential.

The probability in \eqnname~(\ref{eq:pdf}) incorporates and elucidates
commonly used methods for fitting tidal streams. On the one hand are
approaches that search to minimize the spread in energy, integrals of
the motion, or actions (\eg, \citealt{Binney08a,Penarrubia12a},
Sanderson \etal, 2014, in preparation). These approaches ignore the
correlations between actions and angles in the stream and only use
$p(\veco)$; finding the gravitational potential that minimizes the
spread in frequencies (or equivalently, actions) optimizes
$p(\veco)$. Recently, a new method was proposed that only uses the
correlation between the actions and the angles, expressed by the
exponential in \eqnname~(\ref{eq:pdf}) \citep{Sanders13b}; this method
does not use the fact that the spread in actions in a tidal stream is
small (expressed by $p(\veco)$), but only uses the fact that the
angles and actions/frequencies are highly correlated. Approaches that
require stream members to re-unite with the progenitor when
integrating their orbits backward in time
\citep{Johnston99a,PriceWhelan13a} use the full PDF (expressed in
position--velocity space), but cannot be marginalized over unobserved
or noisy phase--space dimensions as easily.

After this paper first appeared, \citet{Sanders14a} presented a
similar tidal-stream model in frequency--angle space as that proposed
here. It employs the algorithm of \citet{Sanders12a} for estimating
the frequencies and angles by approximating the gravitational
potential as a St\"{a}ckel potential in the orbital volume covered by
the stream. The St\"{a}ckel approximation does not perform well for
the eccentric orbits that streams are typically on \citep{Sanders12a}
and the noise induced by the approximation is significant compared to
the internal dispersion around a stream's track, especially for cold
streams. \citet{Sanders14a} focuses on constraining the gravitational
potential using the generative frequency--angle model and demonstrates
that his model, which is essentially the same as that proposed here,
is able to recover the parameters of the potential for a GD-1-like
stream. 

\subsection{Constraining progenitor properties with stream kinematics}

The generative model for a tidal stream proposed here also connects
the dynamics of the stream, in particular its average track and the
dispersion around this track, to the properties of the
progenitor. Primarily, we can constrain $\sigv$, which is proportional
to the velocity dispersion or mass$^{1/3}$ of the progenitor (see
\citealt{Sanders13a}), from the observed width of the stream, even if
the stream's sky projection is the only projection for which we can
resolve the stream. Further $N$-body simulations are necessary to
characterize the exact relation between the \sigv\ parameter and the
velocity dispersion of the progenitor and to determine the importance
of the cluster's internal structure, primarily its concentration, in
determining \sigv.

Our calculation of the mean stream orbit as a function of angle along
the stream also points the way to determining the location of an
``orphan'' stream's progenitor (an orphan stream is a stream for which
the progenitor is not known). This requires high-quality data that
allows the average orbit as a function of position along the stream to
be determined. If this can be done, then the observation of a steady
change in the mean orbit followed by a plateau indicates that the
progenitor lies in the direction of the plateau. The change in the
mean orbit along the stream is a function of the mean orbit offset
($\Delta \Omega^m$ in the direction of $\Delta \opar$), the dispersion
in orbit offsets ($\sigma_{\Omega,1}$ in $\Delta \opar$), and the
disruption time $t_d$. The first two of these can be measured from the
width of the stream and the plateau in the mean orbit as a function of
stream position. The disruption time $t_d$ can then be measured from
the change in mean orbit at the end of the stream. With all of the
parameters of the forward model measured, we can extrapolate the
stream backward to the progenitor position. Similarly, we can
extrapolate a leading stream to its trailing counterpart, if this is
unknown (and vice versa).

While determining the mean orbit as a function of position along the
stream requires high-quality data, this should be achievable in the
near future. We can average or ``stack'' noisily measured phase--space
coordinates in small segments of the stream to determine high quality
average phase--space positions and use these instead of data on
individual stream stars. These considerations show that further
observations of members of the GD-1 and Orphan streams would be highly
informative for determining the location or fate of their progenitors.

\subsection{Baseline models for stream-gap finding}

Besides allowing for sensitive measurements of the Milky Way's
large-scale gravitational potential through the wide arcs traced by
tidal streams, their structure on smaller scales is also sensitive to
more subtle features of the halo's density distribution. In
particular, the perturbations from dark--matter subhalos orbiting
within the Milky Way's halo can have the effect of removing stars from
small segments of a tidal stream \citep{Yoon11a,Carlberg12a}. Thus,
the small-scale density structure of tidal streams can be used to
constrain the subhalo mass function at the high-mass end ($M \gtrsim
10^5\msun$). This holds a great promise for testing the basic
predictions of dark--matter clustering in the $\Lambda$CDM framework
and for shedding light on galaxy formation in the smallest galaxies in
the Universe.

Tidal streams, even in the absence of perturbations from orbiting
dark--matter subhalos, are not entirely smooth as the effects of the
preferential stripping at pericenter and orbital dynamics can create
over- and underdensities along the observed stream track (see
\figurename~\ref{fig:gd1_times}; \citealt{Kuepper10a}). These effects
can lead to spurious detections of gaps in streams that provide a
source of noise in measurements of gaps due to substructure in the
halo \citep{Ngan14a}. 

The generative stream models proposed in this paper can be useful for
generating tailor-made background models for stream gap-finding
algorithms. The generative model allows for the density along the
stream to be predicted quickly for a given progenitor orbit and
distribution of stripping times. The example given in
\figurename~\ref{fig:gd1_pdf_l} using the simple fiducial model shows
that this works well. However, the fiducial model assumes a uniform
distribution of stripping times rather than bursts at pericenter
passages, and therefore it cannot fully model the under- and
over-densities along the stream (\eg, ``feathering'' due to epicyclic
overdensities; see \citealt{Kuepper10a}) in detail. The
mock-stream-generation algorithm given in \sectionname~\ref{sec:mock}
works almost as simply for more complicated models of the stripping
process (\eg, a bursty distribution of stripping times and more
energetic stripping at pericenter; \citealt{Johnston98a}) and it is
therefore straightforward to produce more realistic models of the
small-scale density structure of observed tidal streams. These
background models will allow more sensitive detections of substructure
due to halo substructure.

\section{Conclusion}\label{sec:conclusion}

I have presented a new method for modeling the dynamics of tidal
streams, making extensive use of action--angle variables. This new
framework consists of simple models for the disruption of a star
cluster in frequency--angle space coupled with a novel method for
calculating action--angle variables for any orbit in any
potential. The model is a generative model, meaning that it can be
used to sample mock stream data and evaluate the likelihood of
different models for observed tidal stream data. I have described fast
methods for calculating the mean location of the stream in various
coordinate systems, for estimating the dispersion of the stream and
relating it to the velocity dispersion or mass of the progenitor, and
for marginalizing the stream likelihood over noisily or entirely
unobserved dimensions of phase space.

The framework allows for the proper analysis of data on tidal streams
by taking into account all the relevant dynamical effects, most
notably the stream--single-orbit offset. As such, it should be
immediately useful for the proper analysis of data on the GD-1
\citep{Koposov10a}, Orphan \citep[\eg,][]{Sesar13a}, Pal 5, and other
streams. The ability to quickly marginalize the model over unobserved
dimensions will also prove highly useful when Gaia data for these
streams become available in the near future. The new framework also
provides a straightforward way to generate mock stream data that is
useful for detecting anomalies in the structure of tidal streams, such
as those that can be caused by fly-bys of dark--matter subhalos.

Besides being a practical tool for proper stream fitting, the
framework presented here also clarifies the relation between the
stream, the potential, and properties of the progenitor. I have
discussed how we can relate the properties of a stream to those of its
progenitor, potentially allowing for the unknown progenitor of streams
such as GD-1 or the Orphan stream to be located. Further
investigations of the dynamics of cluster disruption and tidal-stream
generation for different progenitor orbits, a larger and more
realistic set of potentials, and different progenitor structures will
be necessary to perfect the models in frequency--angle space proposed
here and to parameterize the mapping between model parameters and
actual progenitor properties.

The full modeling and action--angle framework presented here is
available as part of the \emph{galpy} Galactic Dynamics
code\footnote{Available at
  \protect{\url{http://github.com/jobovy/galpy}}~.} (J.~Bovy, 2015, in
preparation). A tutorial on how to use this code is given at
\begin{center}
\protect{\small\url{http://galpy.readthedocs.org/en/latest/streamdf.html}}\,.
\end{center}


\acknowledgements It is a pleasure to thank Mark Fardal for comments
on the manuscript and diagnosing problems in the public code that have
significantly improved the paper. I also thank James Binney, David
W. Hogg, Kathryn Johnston, Hans-Walter Rix, Jason Sanders, and Scott
Tremaine for helpful discussions and comments. J.B. was supported by
NASA through Hubble Fellowship grant HST-HF-51285.01 from the Space
Telescope Science Institute, which is operated by the Association of
Universities for Research in Astronomy, Incorporated, under NASA
contract NAS5-26555. J.B.  acknowledges support from SFB 881 (A3)
funded by the German Research Foundation DFG.

\appendix

\section{An efficient, general method for calculating action-angle coordinates using orbit integration}\label{sec:aa}

The wide-spread use of action--angle coordinates, in many ways the
natural coordinates for studying the orbital structure of galaxies, in
dynamical modeling has been frustrated by the difficulty in
calculating the transformation to and from these coordinates,
$(\vecx,\vecv) \leftrightarrow (\vecj,\veca)$. The most general method
for calculating the transformation $(\vecj,\veca) \rightarrow
(\vecx,\vecv)$ is provided by torus modeling \citep{McGill90a}. In the
context of modeling observational data, computing the transformation
$(\vecx,\vecv) \rightarrow (\vecj,\veca)$ is more important and the
most general method consists of iteratively inverting the torus
modeling \citep{McMillan08a}, which is computationally
expensive. Recently, progress has been made in computing
$(\vecx,\vecv) \rightarrow (\vecj,\veca)$ for Milky-Way-like
potentials by either implicitly or explicitly approximating the
gravitational potential as a St\"{a}ckel potential
\citep{Binney12a,Sanders12a}, which allows action--angle coordinates
to be computed with percent-level errors for moderately eccentric
orbits. However, these methods break down for orbits with radial
and/or vertical actions of a similar magnitude as the angular
momentum---that is, orbits that feel the gravitational potential over
a volume large enough that the St\"{a}ckel approximation fails. This
is especially problematic for the orbits of stars in tidal streams, as
the progenitors of tidal streams are typically on quite eccentric
orbits \citep[\eg,][]{Sanders13a}.

Recently, \citet{Fox12a} has suggested a new method for calculating
$\vecj(\vecx,\vecv)$ inspired by torus modeling. This method requires
only a single orbit integration and allows the actions to be
calculated accurately and as precisely as desired (that is, a
convergence criterion can be applied and convergence can be
attained). I describe the Fox method briefly here, discuss how it can
be simplified, and then show how to extend it to also calculate the
frequencies and angles: $(\vecx,\vecv) \rightarrow (\veco,\veca)$.

As in torus modeling, the method of \citet{Fox12a} uses an auxiliary
isochrone potential $\Phi^A$ that should be close to the target
potential $\Phi$ for which the action--angle coordinates should be
calculated (a precise definition of ``close'' will be given
below). The discussion here focuses on loop orbits; for box orbits an
auxiliary harmonic oscillator potential can be used. Actions and
angles $(\vecj^A,\veca^A)$ calculated in the auxiliary potential can
be related to those in the target potential $(\vecj,\veca)$ by a
generating function that can be written as \citep{McGill90a}
\begin{equation}
  S(\veca^A,\vecj) = \veca^A\cdot\vecj+2\sum_{\vecn > 0} S_{\vecn}(\vecj)\sin(\vecn\cdot\veca^A)\,,
\end{equation}
where the $\vecn > 0$ sum is over the integer three-dimensional
half-space that excludes the origin and the $S_{\vecn}$ are functions
of the target actions that are to be determined. The canonical
transformation corresponding to this generating function is
\begin{equation}\label{eq:jjt}
  \vecj^A = \frac{\partial S(\veca^A,\vecj)}{\partial \veca^A} = \vecj + 2\sum_{\vecn > 0} \vecn\,S_{\vecn}(\vecj)\cos(\vecn\cdot\veca^A)\,,
\end{equation}
and
\begin{equation}\label{eq:aat}
  \veca = \frac{\partial S(\veca^A,\vecj)}{\partial \vecj} = \veca^A +2\sum_{\vecn > 0} \frac{\partial S_{\vecn}(\vecj)}{\partial \vecj}\,\sin(\vecn\cdot\veca^A)\,.
\end{equation}
For axisymmetric potentials $J_\phi \equiv L_Z$, such that all
$S_\vecn = 0$ when $n_\phi \neq 0 $ in the above equations. For
axisymmetric potentials we therefore only need to consider
$\theta^A_R$ and $\theta^A_Z$.

\begin{figure}[t!]
 \includegraphics[width=0.48\textwidth,clip=]{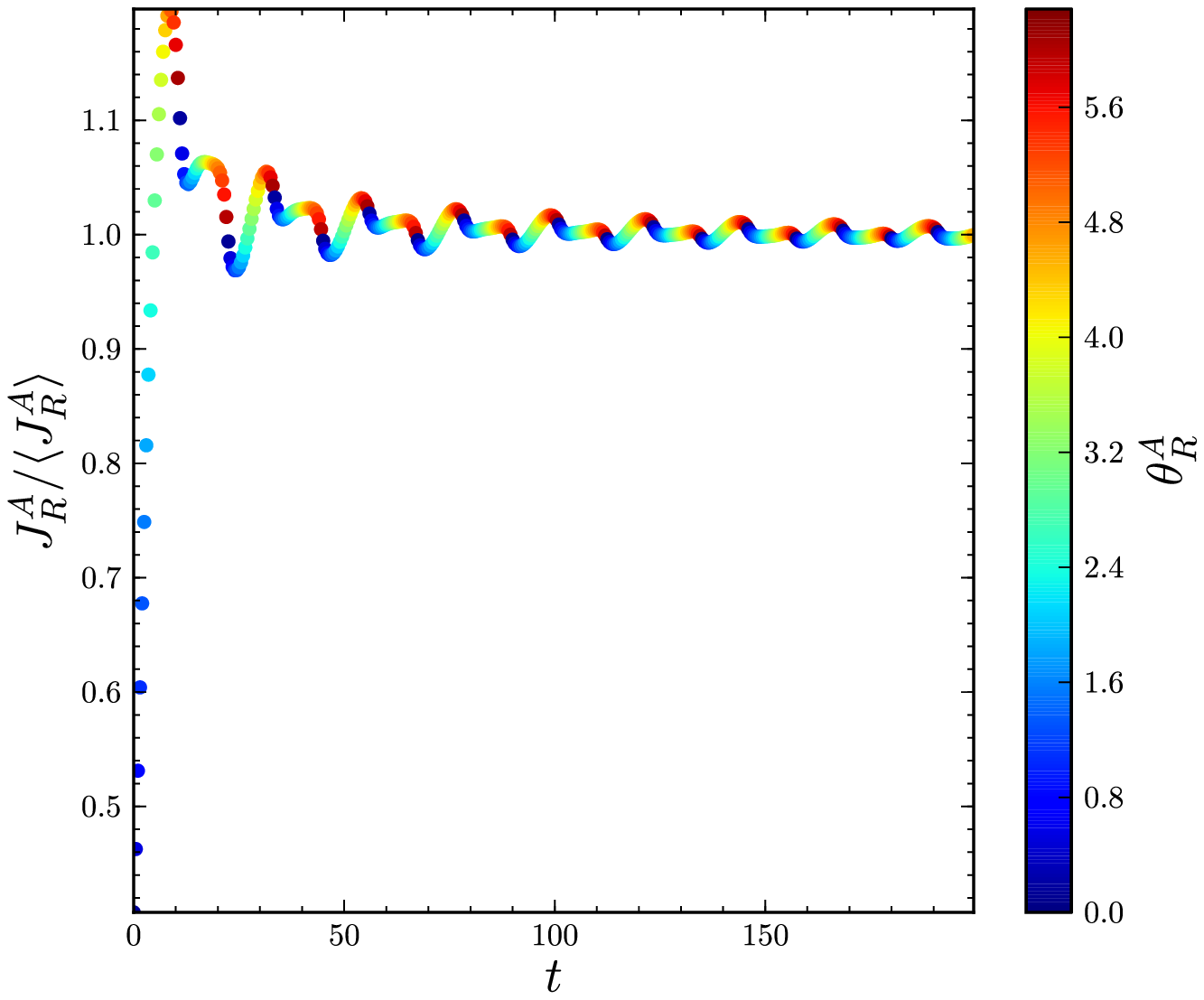}
 \includegraphics[width=0.48\textwidth,clip=]{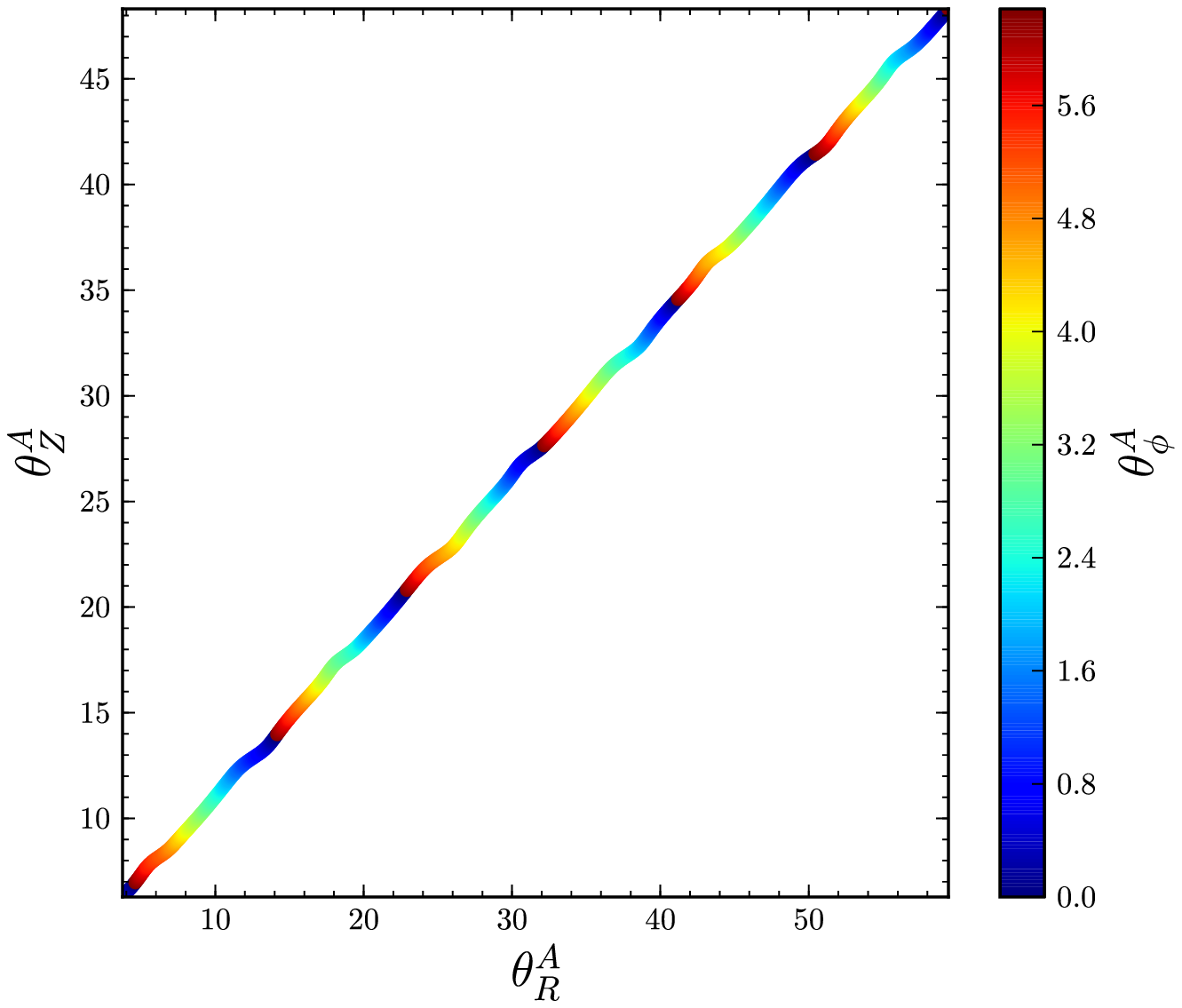}
  \caption{Action--angle calculations using an orbit-integration-based
    method. The left panel shows the cumulative mean of the radial
    actions calculated in an auxiliary isochrone potential calculated
    along the orbit of the progenitor used in the main body of this
    paper. The color-coding corresponds to the radial angle
    (calculated in the isochrone potential) along the orbit. The time
    is in units of $35.56\Myr$ and the radial action is normalized by
    its value at $t=200$. The right panel shows the radial and vertical
    angles calculated in the auxiliary isochrone potential, with
    color-coding corresponding to the azimuthal angle. The wiggles in
    the angles are due to the isochrone potential not being the
    correct potential, but these wiggles are fit and removed to
    calculate the true angles.}\label{fig:aAI}
\end{figure}

The Fox method consists of averaging \eqnname~(\ref{eq:jjt}) over the
three-dimensional torus $\veca^a$ covered by the orbit, such that the
mapping defined by the $S_{\vecn}$ is onto orbital tori of the target
Hamiltonian, the oscillating cosine behavior cancels, and the target
actions are obtained as the auxiliary-angles-averaged auxiliary
actions. Performing these three-dimensional integrals is difficult,
because it is necessary to approximate the three-dimensional volume
$\dd \veca^A$ from a one-dimensional set of points along the
integrated orbit. Here I show that the three-dimensional integrals
over $\veca^A$ can be simplified for non-resonant orbits to
one-dimensional integrals over one of the auxiliary angles along the
path of the orbit. This significantly simplifies the calculations,
especially of $J_\phi$ and $J_Z$ (see below). That is, we integrate
over \eqnname~(\ref{eq:jjt}) as
\begin{equation}\label{eq:jjtint}
  \int \dd \theta_i^A J_i^A = \int \dd \theta_i^A J_i + 2 \sum_{\vecn > 0} \vecn\,S_{\vecn}(J_i)\int \dd \theta_i^A\cos(\vecn\cdot\veca^A)\,,
\end{equation}
where the integral is over auxiliary angles calculated on the path of
the orbit in the target potential. The factor $\vecn\,S_{\vecn}(J_i)$
comes out of the integral as the target actions are conserved along
the orbit. In this Equation, we integrate over the angle coordinate
$\theta_i^A$ associated with the action $J_i$ that we want to compute
(that is, $\theta_R$ for $J_R$ and so on). This is because of the
factor $\vecn$: the condition $\vecn > 0$ means that when calculating
$J_R$ there are terms with $\vecn = (n_R,0,0)$, which would not cancel
according to the argument below when integrating over $\theta_Z$. For
non-resonant orbits and after a long-enough orbit integration, the
angles can be considered to be independent (that is, the orbit ``fills
its torus''), such that we can write (for example, for $\theta_R$)
\begin{equation}
\begin{split}
  \int \dd \theta_R^A\cos(\vecn\cdot\veca^A) & = \mathrm{Re}\left[\int \dd \theta_R^A\,\exp(-i\vecn\cdot\veca^A)\right]\,, \\
  & \approx \mathrm{Re}\left[\exp(-i\,n_\phi\,\theta^A_\phi-i\,n_Z\,\theta^A_Z)\int \dd \theta_R^A\,\exp(-i\,n_R\,\theta^A_R)\right]\,.
\end{split}
\end{equation}
As long as the auxiliary angle $\theta^A_R$ goes through the full range
$0$ to $2\pi$ along the orbit, the oscillatory behavior integrates to
zero and \eqnname~(\ref{eq:jjtint}) simplifies to
\begin{equation}\label{eq:aAIJ}
  J_i = \frac{\int \dd \theta_i^A J_i^A}{\int \dd \theta_i^A}\,,
\end{equation}
that is, the target action is obtained by averaging the auxiliary
action over the auxiliary angle. 

The left panel of \figurename~\ref{fig:aAI} shows a running average of
the radial action calculated using \eqnname~(\ref{eq:aAIJ}) along the
orbit of the progenitor in the simulation used in the main body of
this paper. The time in this figure is in units of $35.56\Myr$, so the
orbit is integrated for a total of $7.11\Gyr$. The auxiliary isochrone
potential has a scale parameter of $6.4\kpc$. After a few radial
periods, the radial action quickly converges with a remaining
oscillatory behavior that is primarily a function of the radial angle,
but that becomes smaller over time. With longer integrations and
shorter time-steps, the radial action can converge to any desired
tolerance. The duration of the orbit integration shown in this figure
is what is typically used in this paper and gives a precision of
$\approx2\,\%$ in the actions.

Given the orbit integration already performed for calculating the
actions, we can use \equationname~(\ref{eq:aat}) to calculate the
frequencies and angles corresponding to the point $(\vecx,\vecv)$
(this procedure is similar to that used in \citealt{McMillan08a}). We
first re-write \equationname~(\ref{eq:aat}) in terms of the angle at
time zero and the frequencies
\begin{equation}\label{eq:aat2}
  \veca = \veca(t=0) + \veco(\vecj) \,t = \veca^A +2\sum_{\vecn > 0} \frac{\partial S_{\vecn}(\vecj)}{\partial \vecj}\,\sin(\vecn\cdot\veca^A)\,.
\end{equation}
This is a linear system of equations for each point along the orbit
with unknowns $\veca(t=0)$, $\veco(\vecj)$, and the functions
$\frac{\partial S_{\vecn}(\vecj)}{\partial \vecj}$. For a finely
integrated orbit, this system is highly over-constrained (at least
when the number of expansion terms is small) and its solution can be
found by simple linear algebra for any desired number of expansion
terms $\sin(\vecn\cdot\veca^A)$. For extra stability, I integrate the
orbit both forward and backward in time, such that the desired angle
$\veca(t=0)$ is at the center of the dependent variable $t$.

The right panel of \figurename~\ref{fig:aAI} shows the auxiliary
radial angle versus the auxiliary vertical angle along the orbit for
the progenitor used in the main body of this paper. The periodic
behavior in the angles is removed before the linear fit described in
the previous paragraph, which is shown in this figure. It is clear
that the contribution of the oscillatory terms
$\sin(\vecn\cdot\veca^A)$ is small, such that a small number of
expansion terms is sufficient (I limit the terms in this axisymmetric
case to $n_R < 4$ and $n_Z < 4$).

The method described here is completely general and the only obstacle
to using it is that a sufficiently close isochrone potential has to be
supplied. Sufficiently close means specifically that the auxiliary
angles have to go through the full range of $0$ to $2\pi$ along the
orbit in the target potential. Whether they do or not is easy to check
and I have found that with limited trial and error a good enough
isochrone potential can be determined quickly for Milky-Way-like
axisymmetric potentials in regions where the flattening is mild (\ie,
in the halo). Whether this method works near the plane of the disk
where the potential is strongly flattened remains to be checked,
although the fact that disk orbits are close to circular makes the
details of the auxiliary potential unimportant. Similarly, the method
should be tested in more detail for triaxial potentials as well. For
analyzing stream data, the necessity of this trial-and-error
determination of the auxiliary potential does not pose a practical
problem, as all of the stream stars are on very similar
orbits. Therefore, the same auxiliary potential can be used for all
action--angle calculations for a given potential and for different
potentials as well, as long as they are not too different. More
widespread use of this technique may require automated methods for
determining a good auxiliary potential. In the unlikely circumstance
that no good auxiliary isochrone potential can be found, the method
discussed here can also be used with other auxiliary potentials for
which the actions and angles can be calculated, such as other
spherical potentials or the family of St\"{a}ckel potentials. This
will, however, add a non-negligible amount of computation for the
necessary numerical integrations.

After this paper appeared, a similar technique for calculating
actions, frequencies, and angles was presented by
\citet{Sanders14b}. Their technique fits for the coefficients of
$\cos(\vecn\cdot\veca^A)$ in \equationname~(\ref{eq:jjt}) to obtain
the actions (similar to how we obtain the frequencies and angles here)
rather than just averaging the auxiliary actions as in
\equationname~(\ref{eq:aAIJ}). They also include a discussion of how
to handle box orbits and a more detailed procedure for finding a good
auxiliary potential. They explicitly and in detail demonstrate that
this new procedure for calculating action--angle coordinates works
well for triaxial potentials.

The action--angle method described in this Appendix is implemented in
the \emph{galpy} Galactic dynamics code.



\begin{thebibliography}{}

\bibitem[{{Binney} \& {Tremaine}(2008)}]{binneytremaine}
  Binney,~J. \& Tremaine,~S. 2008, Galactic Dynamics: Second Edition
\bibitem[Binney(2008)]{Binney08a}
  Binney,~J. 2008, \mnras, 386, L47
\bibitem[Binney(2012)]{Binney12a}
  Binney,~J. 2012, \mnras, 426, 1324
\bibitem[Bovy \etal(2011)]{Bovy11a}
  Bovy,~J., Hogg,~D.~W., \& Roweis,~S.~T. 2011, Ann. Appl. Stat. 5, 1657, arXiv:0905.2979v1
\bibitem[Carlberg(2012)]{Carlberg12a}
Carlberg,~R.~G. 2012, \apj, 748, 20
\bibitem[Dehnen(2000)]{Dehnen00a}
  Dehnen,~W. 2000, \apj, 536, L39
\bibitem[Dehnen(2002)]{Dehnen02a}
  Dehnen,~W. 2002, J.~Comput.~Phys., 179, 27
\bibitem[Eyre \& Binney(2011)]{Eyre11a}
  Eyre,~A. \& Binney,~J. 2011, \mnras, 413, 1852
\bibitem[Fox(2012)]{Fox12a}
  Fox,~M. 2012, M.~Phys.~thesis, Univ.~Oxford, arXiv:1407.1688
\bibitem[Gilks \& Wild(1992)]{Gilks92a}
  Gilks,~W.~R. \& Wild,~P. 1992, Applied Statistics, 41, 337
\bibitem[Helmi \& White(1999)]{Helmi99a}
  Helmi,~A. \& White,~S.~D.~M. 1999, \mnras, 307, 495
\bibitem[Johnston(1998)]{Johnston98a}
  Johnston,~K.~V. 1998, \apj, 495, 297 
\bibitem[Johnston \etal(1999)]{Johnston99a}
  Johnston,~K.~V., Zhao,~H., Spergel,~D.~N., \& Hernquist,~L. 1999, \apj, 512, L109
\bibitem[King(1966)]{King66a}
  King,~I.~R. 1966, \aj, 71, 64
\bibitem[Koposov \etal(2010)]{Koposov10a}
  Koposov,~S.~E., Rix,~H.-W., \& Hogg,~D.~W. 2010, \apj, 712, 260
\bibitem[K\"{u}pper \etal(2010)]{Kuepper10a}
  K\"{u}pper,~A.~H.~W., Kroupa,~P., Baumgardt,~H., \& Heggie,~D.~C. 2010, \mnras, 401, 105
\bibitem[Law et al.(2005)]{Law05a} 
  Law,~D.~R., Johnston,~K.~V., \& Majewski,~S.~R. 2005, \apj, 619, 807
\bibitem[McGill \& Binney(1990)]{McGill90a}
  McGill,~C. \& Binney,~J. 1990, \mnras, 244, 634
\bibitem[McMillan \& Binney(2008)]{McMillan08a}
  McMillan,~P.~J. \& Binney,~J.~J. 2008, \mnras, 390, 429
\bibitem[Ngan \& Carlberg(2014)]{Ngan14a}
  Ngan,~W.~H.~W. \& Carlberg,~R.~G. 2014, \apj, 788, 181
\bibitem[Pe{\~n}arrubia \etal(2012)]{Penarrubia12a}
  Pe{\~n}arrubia, J., Koposov,~S.~E., \& Walker,~M.~G. 2012, \apj, 760, 2
\bibitem[Perryman \etal(2001)]{Perryman01a}
  Perryman,~M.~A.~C., \etal\ 2001, \aap, 369, 339
\bibitem[Price-Whelan \& Johnston(2013)]{PriceWhelan13a}
  Price-Whelan,~A.~M. \& Johnston,~K.~V.~2013, \apj, 778, L12
\bibitem[Sanders(2012)]{Sanders12a}
  Sanders,~J. 2012, \mnras, 426, 128
\bibitem[Sanders \& Binney(2013a)]{Sanders13a}
  Sanders,~J.~L. \& Binney,~J. 2013a, \mnras, 433, 1813
\bibitem[Sanders \& Binney(2013b)]{Sanders13b}
  Sanders,~J.~L. \& Binney,~J. 2013b, \mnras, 433, 1826
\bibitem[Sanders(2014)]{Sanders14a}
  Sanders, J.~L. 2014, \mnras, 443, 423 
\bibitem[Sanders \& Binney(2014)]{Sanders14b}
  Sanders, J.~L. \& Binney, J.\ 2014, \mnras, 441, 3284 
\bibitem[Sesar \etal(2013)]{Sesar13a}
  Sesar,~B., Grillmair,~C.~J., Cohen,~J.~G., \etal\ 2013, \apj, 776, 26
\bibitem[Shoemake(1985)]{Shoemake85a}
  Shoemake,~K. 1985, ACM SIGGRAPH Computer Graphics 3, 45
\bibitem[Teuben(1995)]{Teuben95a}
  Teuben,~P.~J. 1995, in Astronomical Data Analysis Software and Systems IV, 
  ed.~R.~Shaw, H.~E.~Payne and J.~J.~E.~Hayes., PASP Conf Series 77, 398
\bibitem[Tremaine(1999)]{Tremaine99a}
  Tremaine,~S. 1999, \mnras, 307, 877\
\bibitem[Varghese \etal(2011)]{Varghese11a}
  Varghese,~A., Ibata,~R., \& Lewis,~G.~F. 2011, \mnras, 417, 198
\bibitem[Yoon \etal(2011)]{Yoon11a}
  Yoon,~J.~H., Johnston,~K.~V., \& Hogg,~D.~W. 2011, \apj, 731, 58
\end{thebibliography}
\end{document}